\documentclass[11pt,draftcls,onecolumn]{IEEEtran}
\usepackage{amsfonts,epsfig,amsmath,latexsym,amssymb,amscd,multirow,graphicx,lscape,amsmath,amssymb,bm,pifont,graphicx,amssymb,amsmath}
\usepackage{bbm}

\usepackage{undertilde}
\usepackage[usenames,dvipsnames]{xcolor}
\usepackage{graphicx}
\usepackage{lineno}  
\usepackage[normalem]{ulem} 
\usepackage{dsfont}

\def\argmax{\operatornamewithlimits{arg\,max}}

\newcommand{\model}{{\cal M}}

\usepackage{framed}
\usepackage{setspace}

\newcommand{\Data}{{\bf z}}

\newcommand{\state}{{\bm \theta}}
\newcommand{\State}{{\bm \theta}}

\newcommand{\Normal}{{\cal N}}

\newcommand{\Expec}{\mathbb{E}}

\newcommand{\ESS}{\mathbb{ESS}}
\newcommand{\CESS}{\mathbb{CESS}}

\newcommand{\Kernel}{{\cal K}}
\newcommand{\BackKernel}{{\cal L}}

\newcommand{\Target}{{\pi}}
\newcommand{\CoolSchedule}{\phi}
\newcommand{\UnNorTarget}{{\gamma}}
\newcommand{\NormConst}{Z}
\newcommand{\PropDist}{\eta}

\newcommand{\TestFunction}{\varphi}

\newcommand{\NbIterSMC}{T}

\newcommand{\NbParticles}{N}

\newcommand{\ISWeight}{W}
\newcommand{\NormISWeight}{\widetilde{W}}

\newcommand{\IncreWeight}{{{w}}}
\newcommand{\UnNorIncreWeight}{{w}}

\newcommand{\Model}{{\cal M}}

\newcommand{\AcceptRatio}{\alpha}

\newcommand{\Uniform}{{\cal U}}

\newcommand{\ParsIndexSMC}{m}

\newcommand{\NbMCMCmove}{N_{\text{MCMC}}}

\newcommand{\Block}{{\boldsymbol\varrho}}

\definecolor{orange}{rgb}{1,0.5,0}
\definecolor{violet}{rgb}{0.53, 0.0, 0.69}
\definecolor{pakistangreen}{rgb}{0.0, 0.4, 0.0}

\usepackage{multicol}
\usepackage{ragged2e} 
\makeatletter
\newcommand{\justified}{%
  \rightskip\z@skip%
  \leftskip\z@skip}
\makeatother

\usepackage{stfloats}

\usepackage[font=bf]{subfig}

\RequirePackage{dsfont,mathrsfs}
\usepackage{eqparbox}
\usepackage{algorithmic}
\usepackage{algorithm}

\usepackage{rotating}                    

\usepackage{algorithm}
\usepackage{algorithmic}

\makeatletter

\def\cleardoublepage{\clearpage\if@twoside \ifodd\c@page\else%
  \hbox{}%
  \thispagestyle{empty}
  \newpage%
  \if@twocolumn\hbox{}\newpage\fi\fi\fi}

\makeatother
 
{%

\hrulefill
\vspace*{0.5cm}%
\end{minipage}
}

\usepackage{multirow}
{ \begin{list}%
	{$\bullet$}%
	{\setlength{\labelwidth}{25pt}%
	 \setlength{\leftmargin}{30pt}%
	 \setlength{\itemsep}{\parsep}}}%
{ \end{list} }

\renewcommand{\epsilon}{\varepsilon}

\usepackage{bm}
\usepackage{undertilde}

\newcommand{\TimeIndex}{t}

\newcommand{\Correct}[1]{{{#1}}}

\usepackage{cite}

\usepackage[left=2.2cm,right=2.2cm,top=2.4cm,bottom=2.3cm]{geometry}

\begin{document}
\title{New Perspectives on Multiple Source Localization in Wireless Sensor Networks}
\author{Thi Le Thu Nguyen$^{1,2}$, Fran\c{c}ois~Septier$^{1,2}$, Harizo Rajaona$^{2,3}$, Gareth W.~Peters$^{4}$, Ido Nevat$^{5}$ and Yves Delignon$^{1,2}$
\begin{center}
{\footnotesize {\ 
\textit{
$^{1}$ Institut Mines-T\'el\'ecom / T\'el\'ecom Lille,  Villeneuve d'ascq, France.\\
$^{2}$ CRIStAL UMR CNRS 9189, Villeneuve d'ascq, France.\\
$^{3}$ CEA, DAM, DIF, 91297 Arpajon, France\\
$^{4}$ Department of Statistical Sciences, University College London (UCL), London, UK. \\
$^{5}$ Institute for Infocomm Research, Singapore.\\
} } }
\end{center}
}

\maketitle

\vspace{-2cm} 
\begin{abstract}
%
In this paper we address the challenging problem of multiple source localization in Wireless Sensor Networks (WSN). We develop an efficient statistical algorithm, based on the novel application of Sequential Monte Carlo (SMC) sampler methodology, that is able to deal with an unknown number of sources given quantized data obtained at the fusion center from different sensors with imperfect wireless channels.
We also derive the Posterior Cram\'er-Rao Bound (PCRB) of the source location estimate. The PCRB is used to analyze the accuracy of the proposed SMC sampler algorithm and the impact that quantization has on the accuracy of location estimates of the sources.
Extensive experiments show that the benefits of the proposed scheme in terms of the accuracy of the estimation method
that are required for model selection (i.e., the number of sources) and the estimation of the source characteristics
compared to the classical importance sampling method.

\end{abstract}

\section{Introduction} \label{Introduction}
\IEEEPARstart{W}{ireless} sensor networks (WSN) are composed of a large numbers of low-cost,
low-power, densely distributed, and possibly heterogeneous sensors.
WSN increasingly attract considerable research attention due to the large number of applications, such as environmental monitoring \cite{hart2006environmental}, weather forecasts \cite{rajasegarar2014high,kottas2012spatial,fonseca2012stability,frenchspatio}, surveillance \cite{zhang2014distributed,sohraby2007wireless}, health care \cite{lorincz2004sensor}, structural safety and building monitoring\cite{chintalapudi2006monitoring} and home automation \cite{frenchspatio, akyildiz2002wireless}. We consider WSN which consist of a set of spatially distributed sensors that may have limited resources, such as energy and communication bandwidth. These sensors monitor a spatial physical phenomenon containing some desired attribute (e.g pressure, temperature, concentrations of substance, sound intensity, radiation levels, pollution concentrations, seismic activity etc.) and regularly communicate their observations to a Fusion Centre (FC) in a wireless manner (for example, as in \cite{nevat2013random, nevat2014distributed, fazelrandom, matamoros_estimation,schabenberger2010contemporary, akyildiz2004exploiting, vuran2004spatio}). The FC collects these observations and fuses them in order to reconstruct the signal of interest, based on which effective management actions are made \cite{akyildiz2002wireless}.

In this paper, we study the source localization problem which is one important problem that arises in WSN, see for instance the overviews in \cite{patwari2005locating} and \cite{mao2007wireless}.

\subsection{Existing works on source localization from WSN}

A number of works have addressed different aspects of this source localization problem. For instance in a distributed sensor localization problem the work of \cite{patwari2005locating} studied the problem of sensor localizations and the accuracy of such estimation in ad-hoc WSN based on measurements and statistical model design for WSN measurements such as time of arrival, angle of arrival and received signal strength. The observations utilized to make this inference typically come from a WSN in which there is typically a large number of inexpensive sensors that are densely deployed in a region of interest (ROI). Generally, this makes it possible to accurately perform energy based target localization. Signal intensity measurements are very convenient and economical to localize a target, since no additional sensor functionalities and measurement features, such as direction of arrival (DOA) or time-delay of arrival (TDOA), are required.

Such energy-based methods, based on the fact that the intensity (energy) of the signal attenuates as a function of distance from the source, have been proposed and developed in \cite{LiWong2002,LiHu2003,Sheng2005,Blatt:2006in,NiuVarshney2006,Ozdemir2009,Masazade2010}. More precisely, \cite{LiHu2003} developed a least-square method to perform the task of localization for a single source based on the energy ratios between sensors. This was then extended under a Maximum likelihood (ML) based framework for multiple source localizations in \cite{Sheng2005}. 
In this second work, the proposed method uses acoustic signal energy measurements taken at individual sensors to estimate the locations of multiple acoustic sources. By assuming that the number of acoustic sources is known in advance, their estimation approach involved a combination of a multiresolution search algorithm and the use of the expectation-maximization (EM) algorithm. 

However, in both \cite{LiHu2003,Sheng2005}, analog measurements from sensors are required to estimate the source location. For a typical WSN with limited resources (energy and bandwidth), it is important to limit the communication with the network. Therefore, it is often desirable that only binary or multiple bit quantized data be transmitted from local sensors to the fusion center (processing node). Motivated by such constraints, several papers have more recently  proposed source localization techniques using only quantized data \cite{NiuVarshney2006,Ozdemir2009,Masazade2010}. In \cite{NiuVarshney2006}, a ML based approach has been proposed by using multi-bit ($M$-bit) sensor measurements transmitted to the fusion center. In \cite{Masazade2010}, the authors have also developed for the same problem an alternative solution based on an importance sampler which was utilized to approximate the posterior distribution of the single source given the quantized data. However, in both works, perfect communication channels between sensors and the fusion center are assumed. Usually, in a target localization scenario, a large number of sensors are deployed in some area where a line-of-sight between sensors and the FC cannot be always guaranteed. In \cite{Ozdemir2009}, an extension of the ML-based approach previously derived in \cite{NiuVarshney2006} has been proposed in order to incorporate the imperfect nature of the wireless communication channels.

\vspace{0.2cm}
\subsection{Contribution}

In this paper, we generalize previous source localization works by proposing a localization algorithm for an unknown number of sources given some quantized data obtained at the fusion center from different sensors with imperfect wireless channels. The statistical approach we derive is based on the recent and efficient sampling framework known as Sequential Monte Carlo Samplers (SMC Samplers) \cite{Peters:2005ti,DelMoral2006}, and is able to estimate jointly the unknown number of sources as well as their associated parameters (locations and transmitted powers) by providing all the information included in the approximated posterior distribution. In addition, we also derive the PCRB which provides a theoretical performance limit for the Bayesian estimator of the locations as well as the transmitted powers of the $K$ sources.
We demonstrate that the proposed framework provides significant improvement over classical importance sampling type methods that have been used for a single source context in \cite{Masazade2010} and adapted here for multiple sources.

\section{Problem Formulation}
In this section we first present the system model, and then develop the Bayesian framework for jointly estimating the unknown number of sources as well as their locations and transmitted powers.
\subsection{Wireless Sensor Network System Model}
\label{ProblemFormulationSec}

\begin{figure}[htb]
\centering
\includegraphics[width=0.5\textwidth]{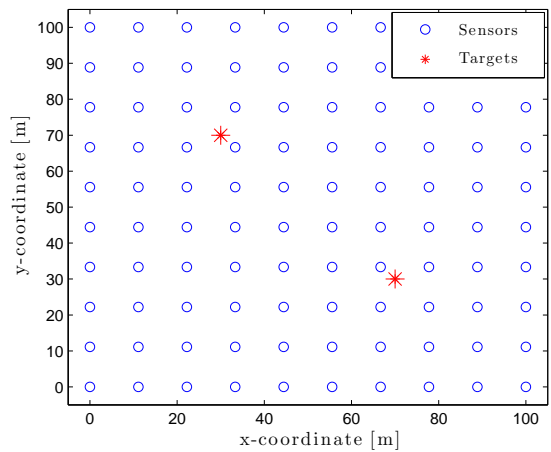}	
\caption{Example of two targets in a grid deployed sensor field.}
\label{fig:Chap3:Scenario}
\end{figure}

As illustrated in Fig. \ref{fig:Chap3:Scenario}, we are interested in localizing an unknown number of targets in a wireless sensor environment where  homogeneous and low-cost wireless sensors are utilized. All the sensors report to a fusion center which then performs the estimation of the target locations based on local sensor observations. Sensors can be deployed in any manner since our approach is capable of handling any kind of deployment as long as the location information for each sensor is available at the fusion center.

Each target is assumed to be a source that follows the power attenuation model. We thus use a signal attenuation model to represent the observed power that is emitted by each target \cite{NiuVarshney2006}. This signal attenuation model is based on the fact that an omnidirectional point source emits signals that attenuate at a rate inversely proportional to the distance from the source, for instance a traveling wave that may be propagating through ground surface or an acoustic pressure wave traveling through free-space medium. 

In this work, as in \cite{Sheng2005}, we will further assume that the intensities of the $K$ sources will be linearly superimposed without any interaction between them. The received signal amplitude at the $i$-th sensor ($i=1,\ldots,N$) is thus given by
\begin{equation}
s_i=a_i+n_i ,
\label{EqGeneralObs}
\end{equation}
where the measurement noise term, $n_i$, is modeled as an additive white Gaussian noise (AWGN), i.e., $n_i\sim \Normal(0,\sigma^2)$ which represents the cumulative effects of sensor background noise and the modeling error of signal parameters (the Gaussian assumption is  generally admitted since the central limit theorem could be applied on a processed signal resulting on the average of the samples received during a time period). The true signal amplitude $a_i$ from all the targets is defined as \cite{Sheng2005}:
\begin{equation}
a_i=\sum_{k=1}^K P_k^{1/2} \left( \frac{d_0}{d_{i,k}} \right)^{\frac{n}{2}} , 
\label{AmplitudeDefLoc}
\end{equation}
where $P_k$ denotes the $k$-th source signal power at a reference distance $d_0$. The signal decay $n$ is approximately $2$ when the detection distance is less than 1km \cite{LiHu2003}. Finally $d_{i,k}$ corresponds to the distance between the $i$-th sensor and the $k$-th target:
\begin{equation}
d_{i,k}=\sqrt{(x_k-c_{x,i})^2+(y_k-c_{y,i})^2} ,
\end{equation}
where $(c_{x,i},c_{y,i})$ and $(x_k,y_k)$ are the coordinates of the $i$-th sensor and the $k$-th target, respectively. In this work, we assume that sensor noises as well as wireless links between the sensors and the fusion center are independent across sensors, and that $\sigma^2$ is known (although it is  not required for our proposed approach to work - this could be indeed embedded in the parameters to be inferred).
	
\begin{figure}[htb]
\centering
\includegraphics[width=12cm]{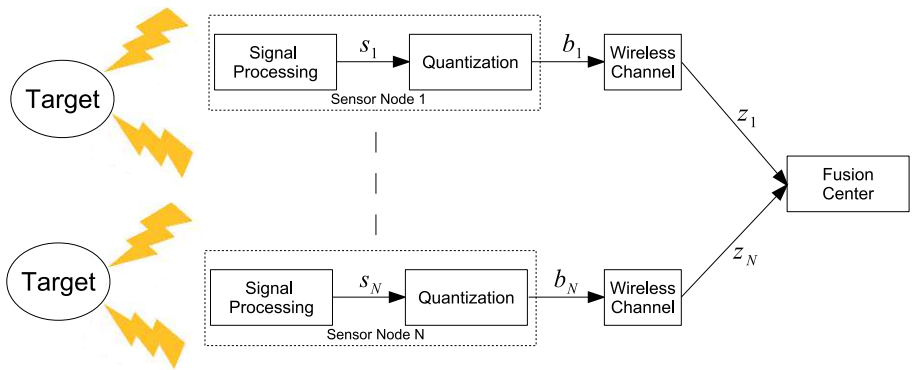}	
\caption{Illustration of the system model.}
\label{fig:Chap3:SystemModel}
\end{figure}	
	
As illustrated in Fig. \ref{fig:Chap3:SystemModel}, at each sensor, the received signal is quantized before being sent to the fusion center. Quantization is done locally at the sensors in order to decrease the communication bandwidth on the sensors thereby reducing energy consumption. The data is quantized using an $M$-bit quantizer ($M\geq 1$) which takes values from 0 to $2^M-1$ where $L=2^M$ is the number of quantization levels. The quantizer of the $i$-th sensor transforms its input $s_i$ to its output $b_i$ through a mapping: $\mathbb{R}\mapsto\left\{0,\ldots,L-1\right\}$ such that
\begin{equation}
b_i=\left\{\begin{array}{ll}
0 & \lambda_{i,0} \leq s_i < \lambda_{i,1}, \\
1 & \lambda_{i,1} \leq s_i < \lambda_{i,2},\\
\vdots & \hspace*{1cm}  \vdots\\
L-1 & \lambda_{i,L-1} \leq s_i < \lambda_{i,L},
\end{array}\right.
\label{QuantizationChap3}
\end{equation}
with $\lambda_{i,0}=-\infty$ and $\lambda_{i,L}=+\infty$. Let $\state_K=\begin{bmatrix}
	P_1,x_1,y_1,\ldots,P_K,x_K,y_K \end{bmatrix}^T$ denote all the $K$ source locations and their associated transmitted powers. Under Gaussian assumption of the measurement noise, the probability that $b_i$ takes a specific value $l\in \{0,\ldots,L-1 \}$ is:
\begin{equation}
p(b_i=l |\state_K)= Q\left( \frac{\lambda_{i,l}-a_i}{\sigma} \right) -Q\left( \frac{\lambda_{i,l+1}-a_i}{\sigma} \right),
\end{equation}
where $Q(\cdot)$ is the complementary distribution function of the Gaussian distribution defined as:
\begin{equation}
Q(x)=\int_{x}^{+\infty} \frac{1}{\sqrt{2\pi}} e^{-\frac{t^2}{2}} dt.
\end{equation}
Finally, the quantized observation are transmitted to the fusion center through an imperfect channel which may introduce transmission errors. Let ${\bm z}=\begin{bmatrix}
	z_1,\ldots,z_N
	\end{bmatrix}$ denote the observations collected at the fusion center via independent channels from the $N$ sensors. As in \cite{nevat2013random, Ozdemir2009,Nevat2014}, the probability of a received observation $z_i$ taking a specific value $j$, given the targets' parameters, $\state_K$, can be written as:
\begin{eqnarray}
p(z_i=j |\state_K)  =  \sum_{m=0}^{L-1} p(z_i=j | b_i=m) p(b_i=m | \state_K) ,
\label{LikelihoodTargetSingleObs}
\end{eqnarray}
where $ p_{j,m}:=p(z_i=j | b_i=m)  $ represents the transition probabilities of the wireless channel, see \cite{nevat2013random, Ozdemir2009,Nevat2014}.

Since sensor noises and wireless links are assumed to be independent, the likelihood function at the fusion center can be written as:
\begin{eqnarray}
p({\bm z}|\state_K) &= &\prod_{i=1}^N p(z_i|\state_K)  \nonumber \\
&= &  \prod_{i=1}^N  \left[ \sum_{m=0}^{L-1} p(z_i | b_i=m) p(b_i=m | \state_K) \right] .
\label{CompleteLikelihood}
\end{eqnarray}
	
Concerning the prior information related to the parameters of interest $\state$, we use in this work:
\begin{eqnarray}
p(\state_K)=\prod_{k=1}^K p(x_k,y_k) p(P_k) ,
\label{PriorTargetChap3}
\end{eqnarray}
where
\begin{align}
\begin{split}
p(x_k,y_k)&=\Normal( {\bm \mu}_p, {\bm \Sigma}_p),\\
p(P_k) & =  {\cal IG} (a,b),
\end{split}
\label{PriorTargetChap3Details}
\end{align}
with ${\bm \mu}_p$ set as the center of the ROI and $ {\bm \Sigma}_p=\text{diag} (\left[\sigma_{p,x}^2 ~ \sigma_{p,x}^2\right])$ is the covariance matrix which is very coarse so that its 99\% confidence region covers the entire ROI. ${\cal IG}(a,b)$ corresponds to the inverse gamma distribution with  $a$ and $b$ being the shape and the scale parameter, respectively. Note that the proposed inference algorithm does not require the prior distributions to be Gaussian and inverse-gamma and will work with other prior distribution choices as required for a given application.

\subsection{Multiple Source Localization in a Bayesian Framework}
In this work, we are interested in estimating the unknown number of sources as well as their parameters (locations and transmitted powers). This problem can therefore be seen as a joint model selection and parameter estimation task. We have a collection of $K_{\max}$ competing models $\{\mathcal{M}_k \}_{k\in\{1,\ldots,K_{\max}\}}$ (corresponding in our case to the number of sources in the ROI) and one of them generates the observations obtained at the fusion center. Associated with each model, there is a vector of parameters $\state_k\in \Theta_k$, where $\Theta_k$ denotes the parameter space of the model ${\cal M}_k$. The objective is to identify the true model as well as to estimate the parameters, $\state_k=\begin{bmatrix} P_1,x_1,y_1,\ldots,P_k,x_k,y_k \end{bmatrix}^T$, associated with this model.

Bayesian inference proceeds from a prior distribution over the collection of models, $p({\cal M}_k)$, a prior distribution for the parameters of each model, $p(\state_k|{\cal M}_k)$ and the likelihood under each model $p({\bm z}|\state_k,{\cal M}_k)$. In order to solve this joint model selection and parameter estimation under this Bayesian framework, we first employ a Maximum A-Posterior (MAP) model selection rule and therefore provides a parameter estimate under the selected model. Following the model selection step is the inference step of the model parameters which can then be deduced from the posterior distribution associated with the model ${\cal M}_{k^*}$, i.e. $p(\state_{k^*}|{\bm z},{\cal M}_{k^*})$.
This procedure is summarised as follows:

\begin{enumerate}
	\item Model selection:
	\begin{align}
\begin{split}
k^*&= \argmax_k \left\{ p({\cal M}_k|{\bm z}) \right\}  \\
&= \argmax_k \left\{p({\bm z}|{\cal M}_k)p({\cal M}_k) \right\} ,
\end{split}
	\label{PosteriorModelChap3}
\end{align}
\item Model parameters estimation via Bayes rule:
\begin{equation}
p(\state_{k^*}|{\bm z},{\cal M}_{k^*})=\frac{p({\bm z}|\state_{k^*},{\cal M}_{k^*}) p(\state_{k^*}|{\cal M}_{k^*})}{p({\bm z}|{\cal M}_{k^*}) }.
\label{NormalizingConstantPosteriorProblem}
\end{equation}
\end{enumerate}
Deriving the expressions in (\ref{PosteriorModelChap3}-\ref{NormalizingConstantPosteriorProblem}) involves calculating the evidence of the $k$-th model ${\cal M}_k$:
\begin{equation}
\begin{split}
\label{EvidenceModel}
p({\bm z}|{\cal M}_k) &= \int_{\Theta_k} p({\bm z}|\state_k,{\cal M}_k) p(\state_k|{\cal M}_k) d\state_k\\
&= \int_{\Theta_k} \prod_{i=1}^N  \left[ \sum_{m=0}^{L-1} p(z_i | b_i=m) p(b_i=m | \state_k) \right] 
\prod_{n=1}^k p(x_n,y_n|{\cal M}_k) p(P_n|{\cal M}_k) d\state_k \\
&= \int_{\Theta_k} 
\prod_{i=1}^N  \left[ \sum_{m=0}^{L-1} p_{i,m} 
 \left(Q\left( \frac{\lambda_{i,m}-a_i}{\sigma} \right) -Q\left( \frac{\lambda_{i,m+1}-a_i}{\sigma} \right) 
\right)\right]
\prod_{n=1}^k  \Normal\left(\begin{bmatrix}x_n\\y_n\end{bmatrix}; {\bm \mu}_p, {\bm \Sigma}_p\right){\cal IG} (P_n;a,b) d\state_k
\end{split}
\end{equation}

Unfortunately, owing to the highly nonlinear observation function of the parameters of interest in Equations (\ref{EqGeneralObs}-\ref{AmplitudeDefLoc}), the integral in (\ref{EvidenceModel}) is intractable. As a result, $\forall k\in \{ 1,\ldots,K_{\max}\}$, both the evidence $p({\bm z}|{\cal M}_k)$ and the posterior distribution of the parameters $p(\state_{k}|{\bm z},{\cal M}_{k})$ are intractable. In this work, we propose to use SMC sampler in order to have an accurate approximation of both quantities.
\section{Proposed Bayesian Algorithm to Multiple Source Localization in WSN}

In this section we first introduce the general principle of SMC samplers, then develop the SMC sampler for multiple source localisation, and finally we derive the point estimate for the parameters of interest.
\subsection{General Principle of SMC Samplers}

Sequential Monte Carlo (SMC) methods are a class of sampling algorithms which combine importance sampling and resampling. They have been primarily used as ``particle filters'' to solve optimal filtering problems; see, for example, \cite{Cappe2007} and \cite{Doucet2009} for recent reviews. In this context, SMC methods/particle filters have enjoyed wide-spread use in various applications (tracking, computer vision, digital communications) due to the fact that they provide a simple way of approximating complex filtering distribution sequentially in time. However in \cite{Peters:2005ti,DelMoral2006}, there have been a range of developments to create a general framework that allows SMC to be used to simulate from a single and static target distribution, thus becoming an interesting alternative to standard MCMC methods as well as to population-based MCMC algorithms. Finally, let us note that there exists a few other SMC methods appropriate for static inference such as annealed importance sampling \cite{Neal2001}, the sequential particle filter of \cite{Chopin2002} and population Monte Carlo \cite{Cappe2004} but all of these methods can be regarded as  a special case of the SMC sampler framework.

The SMC sampler is based on two main ideas:
\begin{enumerate}
\item[a)] Rather than sampling directly the complex distribution of interest, a sequence of intermediate target distributions, $\left\{\Target _{\TimeIndex }\right\}_{\TimeIndex=1}^\NbIterSMC$, is designed, that transitions smoothly from a simpler distribution to the one of interest.
In Bayesian inference problems, the target distribution is the posterior $\Target_{T}(\state  )=p(\state|{\bm z})$, thus a natural choice for such a sequence of intermediate distributions is  to select the following \cite{Neal2001}
\begin{equation}\label{SequenceSMCsampler}
\Target _{\TimeIndex }(\state  ) = \frac{\UnNorTarget_t(\state)}{\NormConst_t}\varpropto p(\state  ) p (\Data | \state  )^{\CoolSchedule _{\TimeIndex }}
\end{equation} 
where $\left \{   \CoolSchedule _{\TimeIndex }\right\} $ is a non-decreasing temperature schedule with $\CoolSchedule _{0}=0$ and $\CoolSchedule _{\NbIterSMC }=1$ and $\UnNorTarget_t(\state)$ corresponds to the unnormalized target distribution $\left( \text{i.e. } \UnNorTarget_t(\state)=p(\state  ) p (\Data | \state  )^{\CoolSchedule _{\TimeIndex }}\right)$ and $\NormConst_t = \int_{\Theta}p(\state  )  p (\Data | \state  )^{\CoolSchedule _{\TimeIndex }} d\state$ is the normalization constant. We  initially target  the prior distribution  $\Target _{0}= p(\state  )$  which is generally easy to sample directly from  and then introduce the effect of the likelihood gradually in order to obtain at the end, $\TimeIndex =\NbIterSMC $, the complex posterior distribution of interest $\Target _{T}(\state  )=p(\state  | \Data )$ as target distribution. 
\item [b)] The idea is to  transform this problem in the standard SMC filtering framework, where the sequence of target distributions on the path-space, denoted by $\{ \widetilde{\Target}_t \}_{t =1}^T$, which admits $\Target_t(x_t)$ as marginals, is defined on the product space, i.e., $\text{supp}(\widetilde{\Target}_t) = \Theta \times \Theta \times ... \times \Theta = \Theta^t$. This novel sequence of joint target distributions $ \tilde{\Target }_{\TimeIndex }$ is defined as follows:
		 \begin{equation}\label{Eq_Artificial_TargetDistribution}
	 	 \widetilde{\Target }_{\TimeIndex } (\state _{1:\TimeIndex })=\dfrac{\widetilde{\UnNorTarget}_{\TimeIndex } (\state _{1:\TimeIndex }) }{\NormConst _{\TimeIndex }} ,
	 \end{equation}
	 where
	\begin{equation}\label{Eq_Artificial_UnNorTargetDistribution}
		 \widetilde{\UnNorTarget  }_{\TimeIndex } (\state _{1:\TimeIndex })= \UnNorTarget  _{\TimeIndex }(\state_{\TimeIndex } ) \prod\limits _{k=1}^{\TimeIndex -1} \BackKernel _{k}(\state _{k+1},\state _{k}) ,
	\end{equation}
	in which  the artificial kernels introduced $\{\BackKernel _{k}\}_{k=1}^{\TimeIndex -1}$ are called \textit{backward} Markov kernels since $\BackKernel _{\TimeIndex } (\state _{\TimeIndex+1},\state _{\TimeIndex})$ denotes the  probability density of moving back from $\state _{\TimeIndex +1}$ to $\state _{\TimeIndex }$.  By using such a sequence of extended target distributions $\left \{  \widetilde{\Target }_{\TimeIndex } \right\}_{\TimeIndex =1}^{\NbIterSMC } $ based on the introduction of backward kernels $\{\BackKernel _{k}\}_{k=1}^{\TimeIndex -1}$, sequential importance sampling can thus be utilized in the same manner as standard SMC filtering algorithms.
\end{enumerate}

	Within this framework, one may then work with the constructed sequence of distributions, $\widetilde{ \Target }_{\TimeIndex }$, under the standard SMC algorithm \cite{Doucet2001}. In summary, the SMC sampler algorithm therefore involves three stages:
\begin{enumerate}
\item{\underline{{\it Mutation:}}, where the particles are moved from $\State _{\TimeIndex -1}$ to $\State _{\TimeIndex }$ via a \textit{mutation kernel} $\Kernel _{\TimeIndex }(\state  _{\TimeIndex -1},\state_ {\TimeIndex })$ also called \textit{forward kernel};}
\item{\underline{{\it Correction:}},  where the particles are reweighted with respect to $\Target _{\TimeIndex }$ via the incremental importance weight (Equation (\ref{Eq_SMC_IncreWeights})); and}
\item{ \underline{{\it Selection:}}, where according to some measure of particle diversity, \Correct{such as} effective sample size, the weighted particles may be resampled in order to reduce the variability of the importance weights. }
\end{enumerate}

	In more detail, suppose that at time $\TimeIndex -1$,  we have a set of weighted particles $\left \{  \State_{1:\TimeIndex-1 }^{(\ParsIndexSMC )} ,\NormISWeight   _{\TimeIndex -1}^{(\ParsIndexSMC )} \right\} _{\ParsIndexSMC =1}^{\NbParticles }$ that approximates  $\tilde{\Target }_{\TimeIndex -1}$ via the empirical measure
	
	\begin{equation}\label{Eq_Approx_Artificial_TargetDistribution}
	{\widetilde{\Target }}_{\TimeIndex-1}^N (d \state _{1:\TimeIndex -1})=\sum\limits _{\ParsIndexSMC =1}^{\NbParticles } \NormISWeight   _{\TimeIndex -1}^{(\ParsIndexSMC )} \delta_{\State _{1:\TimeIndex -1}^{(\ParsIndexSMC )}} (d \state _{1:\TimeIndex -1})
	\end{equation}
	These particles are first propagated to the next distribution $\tilde{\Target }_{\TimeIndex }$ using \Correct{a} Markov kernel $\Kernel _{\TimeIndex } (\state_{\TimeIndex -1},\state _{\TimeIndex })$ to obtain the set of particles $\left  \{\State_{1:\TimeIndex }^{(\ParsIndexSMC )}  \right\} _{\ParsIndexSMC =1}^{\NbParticles }$. Importance Sampling (IS) is then used to correct for the discrepancy between the sampling distribution $\PropDist_{\TimeIndex }(\State_{1:\TimeIndex })$ defined \Correct{as}
	\begin{equation}\label{Eq_SMC_JointImportanceDist}
		\PropDist_{\TimeIndex } (\state_{1:\TimeIndex }^{(\ParsIndexSMC )})=\PropDist _{1}(\state  _{1}^{(\ParsIndexSMC )}) \prod_{k=2}^{\TimeIndex } \Kernel _{k}(\state  _{\TimeIndex -1}^{(\ParsIndexSMC )},\state  _{\TimeIndex }^{(\ParsIndexSMC )}) ,
	\end{equation}
	and $\widetilde{\Target}_{\TimeIndex}(\State_{1:\TimeIndex })$. In this case the new expression for the unnormalized importance weights is given by
	\begin{equation}\label{Eq_SMC_ImportanceWeights}		
			\ISWeight_{\TimeIndex }^{(\ParsIndexSMC )} \varpropto \frac{ \tilde{\Target}_{\TimeIndex } (\state_{1:\TimeIndex }^{(\ParsIndexSMC )})}{\PropDist_{\TimeIndex } (\state _{1:\TimeIndex }^{(\ParsIndexSMC )} )} =\dfrac{\Target _{\TimeIndex }(\state  _{\TimeIndex }^{(\ParsIndexSMC )}) \prod_{s=1}^{\TimeIndex -1} \BackKernel_{s}(\state  _{s+1}^{(\ParsIndexSMC )},\state  _{s}^{(\ParsIndexSMC )})}{\PropDist _{1}(\state  _{1}^{(\ParsIndexSMC )}) \prod_{k=2}^{\TimeIndex } \Kernel _{k}(\state  _{k -1}^{(\ParsIndexSMC )},\state  _{k }^{(\ParsIndexSMC )})} \varpropto  \IncreWeight _{\TimeIndex } (\State _{\TimeIndex -1}^{(\ParsIndexSMC )},\State _{\TimeIndex }^{(\ParsIndexSMC )}) \ISWeight_{\TimeIndex -1}^{(\ParsIndexSMC )} ,
	\end{equation}
	where $\IncreWeight_{\TimeIndex }$, termed the (unnormalized) \textit{incremental weights}, are calculated as,
		\begin{equation}\label{Eq_SMC_IncreWeights}
		\IncreWeight_ {\TimeIndex } (\State _{\TimeIndex -1}^{(\ParsIndexSMC )},\State _{\TimeIndex }^{(\ParsIndexSMC )})=\dfrac{\UnNorTarget _{\TimeIndex }(\state  _{\TimeIndex }^{(\ParsIndexSMC )}) \BackKernel _{\TimeIndex -1}(\state  _{\TimeIndex }^{(\ParsIndexSMC )},\state  _{\TimeIndex -1}^{(\ParsIndexSMC )})}{\UnNorTarget _{\TimeIndex -1}(\state  _{\TimeIndex -1}^{(\ParsIndexSMC )}) \Kernel _{\TimeIndex }(\state  _{\TimeIndex -1}^{(\ParsIndexSMC )},\state  _{\TimeIndex }^{(\ParsIndexSMC )})} .
	\end{equation}
	\Correct{However, as in the particle filter, since the discrepancy between the target distribution $\tilde{\Target}_{\TimeIndex }$ and the proposal $\PropDist_{\TimeIndex }$ increases with $t$, the variance of the unnormalized importance weights tends therefore to increase as well, leading to a degeneracy of the particle approximation. A common criterion used in practice to check this problem is the effective sample size $\ESS$ which can be computed by:}
	\begin{equation}\label{Eq_SMC_ESS}
		\ESS_{\TimeIndex }=\left [  \sum\limits _{\ParsIndexSMC =1}^{\NbParticles } (\NormISWeight  _{\TimeIndex }^{(\ParsIndexSMC )})^{2} \right] ^{-1}=\dfrac{\left ( \sum\limits _{\ParsIndexSMC =1}^{\NbParticles } \ISWeight  _{\TimeIndex -1}^{(\ParsIndexSMC )} \UnNorIncreWeight _{\TimeIndex } (\State _{\TimeIndex -1}^{(\ParsIndexSMC )},\State _{\TimeIndex }^{(\ParsIndexSMC )}) \right) ^{2}}{\sum\limits _{j=1}^{\NbParticles }\left (   \ISWeight  _{\TimeIndex -1}^{(j)} \right) ^{2} \left (\UnNorIncreWeight _{\TimeIndex } (\State _{\TimeIndex -1}^{(j)},\State _{\TimeIndex }^{(j)})   \right) ^{2}} .
	\end{equation}
	If the degeneracy is too high, \Correct{i.e.,} the $\ESS_{\TimeIndex }$ is below a prespecified threshold, $\overline{\ESS}$, then a resampling step is performed. The particles with low weights are discarded whereas particles with high weights are duplicated. After resampling, the particles are equally weighted. 	To sum up the algorithm proceeds as shown in Algorithm \ref{Algo_SMC}.

		 \begin{algorithm}[h]  
			\caption{Generic SMC Sampler Algorithm}
			\label{Algo_SMC}
			 \begin{algorithmic}[1]
 			 	 \small 
  			 	 \STATE \underline{Initialize particle system} 
   				 \STATE \Correct{Sample $\left \{\State _{1}^{(\ParsIndexSMC )}   \right\}_{\ParsIndexSMC =1}^{\NbParticles } \sim \PropDist _{1} (\cdot)$ and compute  $\NormISWeight  _{1}^{(\ParsIndexSMC )}=\left ( \frac{\UnNorTarget _{1}(\state _{1}^{(\ParsIndexSMC )})}{\PropDist _{1}(\state _{1}^{(\ParsIndexSMC )})}  \right) \left [  \sum_{j =1}^{\NbParticles }\frac{\UnNorTarget _{1}(\state _{1}^{(j )})}{\PropDist _{1}(\state _{1}^{(j )})} \right] ^{-1} $ and do resampling if $\ESS < \overline{\ESS}$}
  				 \FOR {$\TimeIndex =2, \ldots, \NbIterSMC $}
  				 		 \STATE \underline{Mutation:} for each $\ParsIndexSMC =1,\ldots,\NbParticles $ : Sample $\State _{\TimeIndex }^{\ParsIndexSMC } \sim \Kernel _{\TimeIndex }(\state_{\TimeIndex -1}^{(\ParsIndexSMC )};\cdot)$ where $\Kernel _{\TimeIndex }(\cdot;\cdot)$ is a $\Target _{\TimeIndex }(\cdot)$ invariant Markov kernel.
   						\STATE \underline{Computation of the weights:} for each $\ParsIndexSMC =1,\ldots,\NbParticles $ 
  							\vspace{0.1cm} $$\ISWeight _{\TimeIndex }^{(\ParsIndexSMC )}= \NormISWeight     _{\TimeIndex -1}^{(\ParsIndexSMC )} \dfrac{\UnNorTarget _{\TimeIndex }(\state  _{\TimeIndex }^{(\ParsIndexSMC )}) \BackKernel _{\TimeIndex -1}(\state  _{\TimeIndex }^{(\ParsIndexSMC )},\state  _{\TimeIndex -1}^{(\ParsIndexSMC )})}{\UnNorTarget _{\TimeIndex -1}(\state  _{\TimeIndex -1}^{(\ParsIndexSMC )}) \Kernel _{\TimeIndex }(\state  _{\TimeIndex -1}^{(\ParsIndexSMC )},\state  _{\TimeIndex }^{(\ParsIndexSMC )})} $$\vspace{0.1cm}    
   						Normalization of the weights : $\NormISWeight _{\TimeIndex }^{(\ParsIndexSMC )}=\ISWeight _{\TimeIndex }^{(\ParsIndexSMC )}\left [  \sum_{j=1} ^{\NbParticles } \ISWeight _{\TimeIndex }^{(j)}\right] ^{-1}$
  						 \STATE \underline{Selection:} if $\ESS_t<\overline{\ESS}$ then Resample 					
     
  				 \ENDFOR
			\end{algorithmic}
	\end{algorithm}

	Let us mention two interesting estimates from SMC samplers. First, since  $\tilde{\Target }_{\TimeIndex }$ admits $\Target _{\TimeIndex }$ as marginals by construction, for any $1 \leq \TimeIndex \leq \NbIterSMC $ , \Correct{the} SMC sampler provides an estimate of this distribution 
	\begin{equation}\label{Eq_SMC_ApproxTarget}
		{\Target }_{\TimeIndex }^\NbParticles(d\state  )=\sum\limits _{\ParsIndexSMC =1}^{\NbParticles } \NormISWeight   _{\TimeIndex }^{(\ParsIndexSMC )} \delta_{\State _{\TimeIndex }^{(\ParsIndexSMC )}} (d\state  )
	\end{equation}
	and an estimate of any expectations of some integrable function $\TestFunction (\cdot )$ with respect to this distribution by
	\begin{equation}
		\Expec _{{\Target }_{\TimeIndex }^{\NbParticles }}\left[ \TestFunction (\state )\right]=\sum_{\ParsIndexSMC =1}^{\NbParticles } \NormISWeight _{\TimeIndex }^{(\ParsIndexSMC )} \TestFunction (\state _{\TimeIndex }^{(\ParsIndexSMC )}) .
		\label{ExpectationApproxSMCSampler}
	\end{equation}
	Secondly,  the \Correct{estimated ratio} of normalizing constants $\dfrac{\NormConst _{\TimeIndex }}{\NormConst _{\TimeIndex -1}}=\dfrac{\int \UnNorTarget _{\TimeIndex }(\state ) d \state  }{\int \UnNorTarget _{\TimeIndex-1 }(\state ) d \state  }$ is given by
	\begin{equation}\label{Eq_Approx_Ratio_NormalizingConstant}		
		\widehat{\dfrac{\NormConst_{\TimeIndex }}{\NormConst_{\TimeIndex -1}}}=\sum\limits _{\ParsIndexSMC =1}^{\NbParticles } \NormISWeight  _{\TimeIndex -1}^{(\ParsIndexSMC )} \UnNorIncreWeight _{\TimeIndex } (\State _{\TimeIndex -1}^{(\ParsIndexSMC )},\State _{\TimeIndex }^{(\ParsIndexSMC )}) .
	\end{equation}
	Consequently, the estimate of $\dfrac{\NormConst_{\TimeIndex } }{\NormConst_{1 }}$ \Correct{is}
	\begin{equation}\label{Eq_Approx_Ratio_NormalizingConstantZ0Zt}	
		\widehat{\dfrac{\NormConst_{\TimeIndex }}{\NormConst_{1 }}}=\prod\limits _{k=2}^{\TimeIndex} \widehat{\dfrac{\NormConst_{k }}{\NormConst_{k-1 }}}=\prod\limits _{k=2}^{\TimeIndex} \sum\limits _{\ParsIndexSMC =1}^{\NbParticles }  \NormISWeight  _{k -1}^{(\ParsIndexSMC )} \UnNorIncreWeight _{k } (\State _{k -1}^{(\ParsIndexSMC )},\State _{k }^{(\ParsIndexSMC )}) .
	\end{equation}
	If the resampling scheme used is unbiased, then (\ref{Eq_Approx_Ratio_NormalizingConstantZ0Zt}) is also unbiased  whatever the number of particles used \cite{DelMoral2000}. Moreover, the complexity of this algorithm is  $O(\NbParticles )$ and it can be easily  parallelized.

	To conclude this section, let us summarize the advantages of SMC samplers over traditional and population-based MCMC methods. First, unlike MCMC, SMC methods do not require any burn-in period and do not face the sometimes contentious issue of diagnosing convergence of a Markov chain. Secondly, as discussed in \cite{Jasra2007}, compared to population-based MCMC methods, the SMC sampler is a richer class of method since there is substantially more freedom in specifying  the mutation kernels in SMC: kernels \Correct{do not need} to be reversible or even Markov (and hence time adaptive). Finally, unlike MCMC, SMC samplers provide an unbiased estimate of the normalizing constant of the posterior distribution which will be one of the quantities of interest in the inference problem tackled in this paper related to finding the number of targets that are present in the ROI.

\subsection{Proposed SMC Samplers for Bayesian Multiple Source Localization}

Since the evidence of the model ${\cal M}_{k}$ corresponds to the normalizing constant of the posterior distribution of the parameters associated \Correct{with} this model, i.e.:
\begin{equation}
p(\state_{k}|{\bm z},{\cal M}_{k})=\frac{p({\bm z}|\state_k,{\cal M}_k) p(\state_k|{\cal M}_k)}{p({\bm z}|{\cal M}_k)}=\frac{p({\bm z}|\state_k,{\cal M}_k) p(\state_k|{\cal M}_k)}{\int_{\Theta_k} p({\bm z}|\state_k,{\cal M}_k) p(\state_k|{\cal M}_k) d\state_k} ,
\end{equation}
we propose to use  the following procedure:
\begin{enumerate}
\item For each model $\model_k$, $k \in 1,\ldots,K_{\max}$ : approximate the conditional parameter posterior distribution $p(\state_{k}|{\bm z},{\cal M}_{k})$ as well as the marginal likelihood $p({\bm z}|{\cal M}_k)$ using $K_{\max}$ SMC sampler algorithms in parallel (one for each model $\{{\cal M}_k\}_{k\in \{1,\ldots,K_{\max}\}}$) using Equations (\ref{Eq_SMC_ApproxTarget}) and (\ref{Eq_Approx_Ratio_NormalizingConstantZ0Zt}), respectively.
\item Perform the MAP Model selection rule:
\begin{equation}
k^*=  \argmax_k \left\{\hat{p}({\bm z}|{\cal M}_k)p({\cal M}_k) \right\} ,
\label{MAPCriterionModel}
\end{equation}
with $\hat{p}({\bm z}|{\cal M}_k)$ corresponds to the unbiased estimate obtained from the $k$-th SMC sampler.
\item Provide a parameter estimate under the selected model, e.g. the minimum mean square (MMSE) estimate, using the empirical approximation of the posterior distribution $p(\state_{k^*}|{\bm z},{\cal M}_{k^*})$ given by the $k^*$-th SMC sampler.
\end{enumerate}

Let us now describe in more details the different choices required in the design of each SMC sampler: the appropriate sequence of distributions $\{\Target_{\TimeIndex  }\}_{1 \leq \TimeIndex \leq \NbIterSMC }$, the choice of both the mutation kernel $\{\Kernel _{\TimeIndex }\}_{2 \leq \TimeIndex \leq \NbIterSMC }$ and the backward mutation kernel $\{\BackKernel _{\TimeIndex-1 }\}_{\TimeIndex =2}^{\NbIterSMC }$.

\subsubsection{\underline{Sequence of distributions $\Target _{k,\TimeIndex }$}}
\label{ProcedureAdaptiveForward}

An annealing procedure which progressively introduces the effect of the likelihood function is chosen as the sequence of intermediate target distributions, i.e. for the $k$-th SMC sampler dealing with model ${\cal M}_k$:
\begin{equation}\label{SequenceSMCsampler}
\Target _{k,\TimeIndex }(\state_{k,t}  ) = \frac{\UnNorTarget_{k,t}(\state_{k,t})}{\NormConst_t}\varpropto p(\state_{k,t} | {\cal M}_k ) p (\Data | \state_{k,t},{\cal M}_k  )^{\CoolSchedule _{k,\TimeIndex }} ,
\end{equation} 
where $\left \{   \CoolSchedule _{k,\TimeIndex }\right\} $ is a non-decreasing temperature schedule with $\CoolSchedule _{k,0}=0$ and $\CoolSchedule _{k,\NbIterSMC }=1$. The question that arises is how to choose this non-decreasing temperature schedule $ \{\CoolSchedule _{k,\TimeIndex }\}_{t=1,\ldots,T}$. Several statistical approaches have been proposed in order to automatically obtain such a schedule via the optimization of some criteria, which are known as \textit{on-line} schemes.  \cite{jasra2011}  proposed an adaptive selection method based on controlling the rate of the effective sample size ($\ESS_{k,\TimeIndex }$), defined in  (\ref{Eq_SMC_ESS}). 
\Correct{This scheme thus provides an automatic method to obtain the tempering schedule such that the $\ESS$ decays in a regular predefined way.  However, one major drawback of such an approach is that the $\ESS_{k,\TimeIndex }$ of the current sample weights corresponds to some empirical measure of the accumulated discrepancy between the proposal and the target distribution since the last resampling time. As a consequence, it does not really represent the dissimilarity between each pair of successive distributions   unless resampling is conducted after every iteration.
		 
		 In order to handle this problem,  \cite{Yan2013} proposed a slight modification of the $\ESS$, named the \textit{conditional} $\ESS$ ($\CESS$), by considering how good an importance sampling proposal  $\Target _{k,\TimeIndex-1 } $ would be for the estimation of expectation under $\Target _{k,\TimeIndex } $.  At the $t$-th iteration, this quantity  is defined as follows: }
		\begin{equation}\label{Eq_CESS}
			\CESS_{k,\TimeIndex } =\left [\sum\limits_{i=1}^{\NbParticles }\NbParticles \NormISWeight_{k,\TimeIndex-1 } ^{(i)} \left (\dfrac{\UnNorIncreWeight _{k,\TimeIndex } ^{(i)}}{\sum_{j = 1}^{\NbParticles }\NbParticles \NormISWeight_{k,\TimeIndex-1 } ^{(j)}\UnNorIncreWeight _{k,\TimeIndex } ^{(j)}}\right )^{2}\right ]^{-1}=\dfrac{\left (\sum_{i=1}^{\NbParticles }\NormISWeight_{k,\TimeIndex-1 } ^{(i)}\UnNorIncreWeight _{k,\TimeIndex } ^{(i)}\right )^{2}}{\sum_{j= 1}^{\NbParticles }\frac{1}{\NbParticles }\NormISWeight_{k,\TimeIndex-1 } ^{(j)}(\UnNorIncreWeight _{k,\TimeIndex } ^{(j)})^{2}} .
		\end{equation}
In this work, this  $\CESS$ proposed in \cite{Yan2013} will be used in all the $K_{\max}$ SMC samplers that are run in parallel for each model in order to have an automatic specification of their individual temperature scheduling process. Owing to the on-line nature of this $\CESS$-based strategy, the total number of iterations performed by each sampler is not fixed and does not required to be specified prior to the simulation.

\subsubsection{\underline{Sequence of mutation kernels $\Kernel _{k,\TimeIndex }$}}

	The performance of SMC sampler \Correct{depends} heavily upon the selection of the transition kernels $\left \{  \Kernel _{k,\TimeIndex } \right\} _{\TimeIndex =2}^{\NbIterSMC }$  and the auxiliary backward kernels $\left \{ \BackKernel _{k,\TimeIndex -1}  \right\}_{\TimeIndex =2}^{\NbIterSMC } $. 
		  There are many possible choices for $\Kernel _{k,\TimeIndex }$ which have been discussed in \cite{Peters:2005ti,DelMoral2006}. In this study, we propose to employ  MCMC kernels of invariant distribution $\Target _{k,\TimeIndex }$ for $\Kernel _{k,\TimeIndex }$. This is an attractive strategy since we can use the vast literature on the design of efficient MCMC algorithms to build effective and efficient importance distributions  \cite{Robert2004}.
		  
		  More precisely, in this work, since we are interested in a complex model with potentially high-dimensional and multimodal posterior distribution, a series  Metropolis-within-Gibbs kernels with local moves \cite{Robert2004} will be employed in order to successively move the $B_k$ sub-blocks of the state of interest, $\State_{k,t} =[\Block _{k,t,1}, \Block _{k,t,2}, \cdots, \Block _{k,t,B_k}]$. In this work, the sub-block corresponds to the parameters associated to one target, i.e. $\Block _{k,t,b}=\begin{bmatrix} P_b,x_b,y_b \end{bmatrix}^T$ for $b=1,\ldots,B_k=k$. A random walk proposal distribution is used for each sub-block with a multivariate Gaussian distribution as proposal at the $i$-th iteration of the forward kernel:
\begin{equation}
\Block _{k,t,b}^{*}=\Block_{k,t,b}^{i-1} + {\bm \epsilon}_b^i,
\end{equation}		  
\Correct{in which $ {\bm \epsilon}_{b}^i$ is a Gaussian random variable with zero mean and $3\times3$ covariance matrix  ${\bm \Sigma}$}.  The Metropolis within Gibbs used in the implementation of the SMC sampler in this  paper is summarized in Algorithm \ref{Algo_AMWG}.

 	   \begin{algorithm}[h]  
			\caption{Metropolis-within-Gibbs Kernel $\Kernel _{k,\TimeIndex }(\cdot;\cdot)$ for the $m$-th particle}
			\label{Algo_AMWG}
			 \begin{algorithmic}[1]
 			 	 \small 
  				\STATE \underline{Initialization} Set $[\Block _{k,t,1}^0, \ldots, \Block _{k,t,k}^0]=\State_{k,t-1}$
  				 \FOR {$i =1, \ldots, \NbMCMCmove $}
  				 \FOR{$b=1,\ldots,k$}
					\STATE Sample $\Block _{k,t,b}^* \sim {\cal N} \left( \Block_{k,t,b}^{i-1}, {\bm \Sigma} \right)$ 
					\STATE Compute the Acceptance ratio:
					$$\AcceptRatio (\state_k^{*},\state_k)=\min \left \{ 1,\dfrac{p(\Data | \state_k^{*} ,\Model_k )^{\CoolSchedule _{k,\TimeIndex }} p(\state_k^{*} )}{p(\Data | \state_k ,\Model_k )^{\CoolSchedule _{k,\TimeIndex }} p(\state_k )}\right\} $$
					with $\state_k^{*} =[\Block _{k,t,1}^{i}, \ldots,\Block _{k,t,b-1}^{i}, \Block _{b}^{*},\Block _{k,t,b+1}^{i-1},\ldots,\Block _{k,t,k}^{i-1}]$
					and $\state_k =[\Block _{k,t,1}^{i}, \ldots,\Block _{k,t,b-1}^{i}, \Block _{b}^{i-1},\Block _{k,t,b+1}^{i-1},\ldots,\Block _{k,t,k}^{i-1}]$
			\STATE Sample random variate $u$ from $\Uniform (0,1)$   					
					\IF{$u \leq \AcceptRatio (\State_k ^{*},\State_k)$}
						\STATE $\Block_{k,t,b}^{i}=\Block_{k,t,b} ^{*}$	
					\ELSE
						\STATE $\Block_{k,t,b}^{i}=\Block_{k,t,b}^{i-1}$
					\ENDIF			
					\ENDFOR		
  				 \ENDFOR
  				\STATE  Set the new particle value at time $t$ as $\State_{k,t}^{(m)} =[\Block_{k,t,1}^{\NbMCMCmove}, \ldots, \Block_{k,t,k}^{\NbMCMCmove}]$
			\end{algorithmic}
	\end{algorithm}

	\subsubsection{\underline{Sequence of  backward kernels $\BackKernel _{k,\TimeIndex }$}}

		The backward kernel $\BackKernel _{k,\TimeIndex }$ is arbitrary, however as discussed in \cite{DelMoral2006}, it should be optimized with respect to mutation kernel $\Kernel _{k,\TimeIndex }$ to obtain good performance.  In \cite{Peters:2005ti,DelMoral2006}, it was established that the backward kernel which minimizes the variance of the unnormalized importance weights, $\ISWeight _{k,\TimeIndex }$, are given by
	  \begin{equation}\label{Eq_Optimal_BackwardKernel}
	  	\BackKernel _{k,\TimeIndex }^{\text{opt}} (\state  _{k,\TimeIndex +1},\state  _{k,\TimeIndex })=\dfrac{\PropDist _{k,\TimeIndex }(\state  _{k,\TimeIndex }) \Kernel _{k,\TimeIndex +1}(\state  _{k,\TimeIndex },\state  _{k,\TimeIndex +1})}{\PropDist _{k,\TimeIndex +1}(\state  _{k,\TimeIndex +1})} .
	  \end{equation}
	  However, it is typically impossible to use these optimal kernels  as they rely on marginal distributions $\PropDist _{k,\TimeIndex }(\state  _{k,\TimeIndex })$  which do not admit any closed form expression, especially if an MCMC kernel is used as $\Kernel _{k,\TimeIndex }$ which has a  $\Target _{k,\TimeIndex }$-invariant distribution. Thus we can either choose to approximate $\BackKernel _{k,\TimeIndex }^{\text{opt}}$ or choose kernels $\BackKernel _{k,\TimeIndex }$ so that the importance weights are easily calculated or have a familiar form. If an MCMC kernel is used as forward mutation kernel, the following $\BackKernel _{k,\TimeIndex }$ is employed 
	\begin{equation}\label{Eq_SubOptimal_BackwardKernel}		
		\BackKernel _{k,\TimeIndex -1} (\state _{k,\TimeIndex },\state _{k,\TimeIndex -1})=\dfrac{\Target _{k,\TimeIndex }(\state _{k,\TimeIndex -1}) \Kernel _{k,\TimeIndex }(\state _{k,\TimeIndex -1},\state _{k,\TimeIndex })}{\Target _{k,\TimeIndex }(\state _{k,\TimeIndex})} ,
	\end{equation}
which is a good approximation of the optimal backward kernel if the discrepancy between $\Target _{k,\TimeIndex }$ and $\Target _{k,\TimeIndex -1}$ is small; note that (\ref{Eq_SubOptimal_BackwardKernel}) is the reversal Markov kernel associated with $\Kernel_{k,\TimeIndex }$. In this case,  the unnormalized incremental weights becomes for the SMC sampler associated to the $k$-th model becomes
	\begin{equation}\label{Eq_UnNormalized_IncrementWeights1}
		\UnNorIncreWeight _{k,\TimeIndex }^{(\ParsIndexSMC )} (\State _{k,\TimeIndex -1}^{(\ParsIndexSMC )},\State _{k,\TimeIndex }^{(\ParsIndexSMC )})=\dfrac{\UnNorTarget_{k,\TimeIndex }(\state  _{k,\TimeIndex -1}^{(\ParsIndexSMC )})}{\UnNorTarget_{k,\TimeIndex-1 }(\state  _{k,\TimeIndex -1}^{(\ParsIndexSMC )})}=  p(\Data | \state _{k,\TimeIndex -1}^{(\ParsIndexSMC )} ,{\cal M}_k )^{(\CoolSchedule _{k,\TimeIndex }-\CoolSchedule _{k,\TimeIndex -1})}
	\end{equation}	
where $p(\Data | \state _{k,\TimeIndex -1}^{(\ParsIndexSMC )} ,{\cal M}_k )$ is defined in Eq. (\ref{CompleteLikelihood}). Expression (\ref{Eq_UnNormalized_IncrementWeights1}) is remarkably easy to compute and valid regardless of the MCMC \Correct{kernel} adopted. Note that $\CoolSchedule _{k,\TimeIndex }-\CoolSchedule _{k,\TimeIndex -1}$ is the step length of the cooling schedule of the likelihood at time $\TimeIndex $ for the $k$-th sampler. As this step becomes larger, the discrepancy between $\Target _{k,\TimeIndex }$ and $\Target _{k,\TimeIndex -1}$ increases, leading to an increase in the variance of the importance sampling approximation. Thus, it is important to construct a smooth sequence of distributions $\left \{ \Target _{k,\TimeIndex }  \right\} _{0 \leq \TimeIndex \leq \NbIterSMC }$ by judicious choice of an associated real sequence $\left \{ \CoolSchedule _{k,\TimeIndex }  \right\} _{\TimeIndex =0}^{\NbIterSMC }$ as discussed in the previous section.
	
	Let us remark that when such backward kernel is used, the unnormalized incremental weights in (\ref{Eq_UnNormalized_IncrementWeights1}) at time $t$ does not depend on the particle value at time $t$ but just on the previous particle set. It is known that in such cases, the particles $\left\{ \state _{k,\TimeIndex }^{(\ParsIndexSMC )}   \right\}$ should be sampled after the weights 	$\left\{\ISWeight_{k,\TimeIndex }^{(\ParsIndexSMC )}\right\}$ have been computed and after the particle approximation  $\left\{ \NormISWeight_{k,\TimeIndex }^{(\ParsIndexSMC )},\state _{k,\TimeIndex-1 }^{(\ParsIndexSMC )}   \right\}$ has possibly been resampled.

	\vspace*{0.3cm}
	
	Based on this discussion regarding the different choices, the SMC sampler used in this paper is summarized in Algorithm \ref{Algo_SMCSpecific}.

 \begin{algorithm}[h]  
			\caption{$k$-th SMC Sampler Algorithm targeting the posterior distribution $p( \state _{k}|\Data  ,{\cal M}_k )$}
			\label{Algo_SMCSpecific}
			 \begin{algorithmic}[1]
 			 	 \small 
  			 	 \STATE \underline{Initialize particle system} (set $\TimeIndex=1$ and $\CoolSchedule _{k,1 }=0$)
   				 \STATE Sample $\left \{\State _{k,1}^{(\ParsIndexSMC )}   \right\}_{\ParsIndexSMC =1}^{\NbParticles } \sim p(\State _{k}|{\cal M}_k)$ and set  $\NormISWeight  _{k,1}^{(\ParsIndexSMC )}=1/N$ 
  				 \WHILE{$\CoolSchedule _{k,\TimeIndex }<1 $}
  				 \STATE $\TimeIndex=\TimeIndex+1$
  				 \STATE \underline{Automatic Cooling procedure:} Use a binary search to find $\phi^*$ such that $\CESS_{k,t}^{\phi^*} = \overline{\CESS}$ and set $\CoolSchedule _{k,\TimeIndex }=\phi^*$ if $\phi^*<1$ otherwise $\CoolSchedule _{k,\TimeIndex }=1$
					\STATE \underline{Computation of the weights:} for each $\ParsIndexSMC =1,\ldots,\NbParticles $ 
  							\vspace{0.1cm} $$\ISWeight _{k,\TimeIndex }^{(\ParsIndexSMC )}= \NormISWeight     _{k,\TimeIndex -1}^{(\ParsIndexSMC )} p(\Data | \state _{k,\TimeIndex -1}^{(\ParsIndexSMC )} ,{\cal M}_k )^{(\CoolSchedule _{k,\TimeIndex }-\CoolSchedule _{k,\TimeIndex -1})}$$\vspace{0.1cm}    
   						Normalization of the weights : $\NormISWeight _{k,\TimeIndex }^{(\ParsIndexSMC )}=\ISWeight _{k,\TimeIndex }^{(\ParsIndexSMC )}\left [  \sum_{j=1} ^{\NbParticles } \ISWeight _{k,\TimeIndex }^{(j)}\right] ^{-1}$  				 
  				 		 \STATE \underline{Selection:} if $ESS_{k,t}<\overline{\ESS}$ then Resample
  				 		 \STATE \underline{Mutation:} for each $\ParsIndexSMC =1,\ldots,\NbParticles $ : Sample $\State _{k,\TimeIndex }^{(\ParsIndexSMC) } \sim \Kernel _{k,\TimeIndex }(\state_{k,\TimeIndex -1}^{(\ParsIndexSMC )};\cdot)$ where $\Kernel _{k,\TimeIndex }(\cdot;\cdot)$ is a $\Target _{k,\TimeIndex }(\cdot)$ invariant Markov kernel described in more details in Algo. \ref{Algo_AMWG}.
  				 \ENDWHILE
  				 \STATE \underline{Output:} 
  				 \STATE Unbiased approximation of the marginal likelihood : $p(\Data | {\cal M}_k ) \approx \prod\limits _{n=2}^{\TimeIndex} \widehat{\dfrac{\NormConst_{k,n }}{\NormConst_{k,n-1 }}}=\prod\limits _{n=2}^{\TimeIndex} \sum\limits _{\ParsIndexSMC =1}^{\NbParticles }  \NormISWeight  _{k,n -1}^{(\ParsIndexSMC )} \UnNorIncreWeight _{k,n } (\State _{k,n -1}^{(\ParsIndexSMC )},\State _{k,n }^{(\ParsIndexSMC )})$
  				  \STATE Empirical approximation of the posterior distribution 	$p( \state_k|\Data ,{\cal M}_k ) \approx {\Target }_{k,\TimeIndex }^\NbParticles(d\state_k  )=\sum\limits _{\ParsIndexSMC =1}^{\NbParticles } \NormISWeight   _{k,\TimeIndex }^{(\ParsIndexSMC )} \delta_{\State _{k,\TimeIndex }^{(\ParsIndexSMC )}} (d\state_k  )$
			\end{algorithmic}
	\end{algorithm}	

\subsection{Point Estimate for the parameters of interest}
\label{ProposedLabelSwitching}

Once the model has been selected using the MAP criterion described in (\ref{MAPCriterionModel}), the MMSE estimate of the parameters for the $k^*$-th model is obtained using (\ref{ExpectationApproxSMCSampler}):
\begin{equation}
\widehat{\state}_{k^*}=\Expec _{{\Target }_{{k^*},T }^{\NbParticles }}\left[ \TestFunction (\state )\right]=\sum_{\ParsIndexSMC =1}^{\NbParticles } \NormISWeight _{{k^*},T }^{(\ParsIndexSMC )} \TestFunction (\state _{{k^*},T }^{(\ParsIndexSMC )}) ,
\end{equation}
where $T$ denotes the last iteration of the ${k^*}$-th SMC sampler, since in this last iteration, the system of weighted particles represents an empirical approximation of the target posterior distribution, i.e.:
\begin{align}
	p(\state_{{k^*}}|{\bm z},{\cal M}_{{k^*}})={\Target }_{{k^*},T }(\state_{k^*}  )=\sum\limits _{\ParsIndexSMC =1}^{\NbParticles } \NormISWeight   _{{k^*},T }^{(\ParsIndexSMC )} \delta_{\State _{{k^*},T }^{(\ParsIndexSMC )}} (d\state_{k^*}  ) .
\end{align}

Unfortunately, owing to the non-identifiability of the target label in the likelihood and to the same prior for each target, the posterior distribution will be multimodal (as it will be illustrated in Fig \ref{fig:MultimodalPosteriorCase1}). The posterior is indeed invariant under the permutations of source parameters, i.e.,
\begin{equation}
p(\state_{{k^*}}|{\bm z},{\cal M}_{{k^*}})=p(\vartheta(\state_{{k^*}})|{\bm z},{\cal M}_{{k^*}})
\end{equation}
where $\vartheta(\cdot) \in {\cal P}$ denotes any the permutation for which the posterior is invariant and ${\cal P}$ is the set of these permutations.

 In such a case, the MMSE estimate would lead to very poor performance if selected as a point estimate of the source  parameters. The problem of having a Monte-Carlo algorithm that approximates such a multimodal target posterior, which is invariant under permutation, is known in the literature as the \textit{label switching problem} \cite{Stephens2000}.

There exists many algorithms that have been proposed in order to deal with this label switching problem in the class of Monte-Carlo algorithms. A recent and detailed review of these techniques can be found in \cite{BardenetThesis2012}. Here, we are interested in only post-processing technique in order to extract an accurate point-estimate of the state of interest from our particle approximation of the posterior distribution. One of the most commonly used relabeling algorithms is the one proposed in \cite{Stephens2000}.

Let us denote the unweighted set of particles obtained at the last iteration of the SMC sampler that targets the posterior distribution of the selected model by $\utilde{\state}=\left\{\state_{k^*}^{(1)},\ldots,\state_{k^*}^{(N)}\right\}$. In the algorithm proposed in \cite{Stephens2000}, one performs inference tasks (e.g. point estimation) as usual but with the relabeled samples, defined as:
\begin{equation}
{\bm \vartheta}(\utilde{\state})=\left(\vartheta_1(\state^{(1)}),\ldots,\vartheta_N(\state^{(N)})\right) ,
\end{equation}
where
\begin{equation}
{\bm \vartheta}=\left(\vartheta_1,\ldots,\vartheta_N \right)=\argmax_{ {\cal P} \times \cdots \times  {\cal P}} L( \utilde{\state},{\bm \vartheta}) ,
\end{equation}
and $L(\cdot)$ is a user-defined cost-function, which is generally chosen as:
\begin{equation}
L( \utilde{\state},{\bm \vartheta})=\prod_{i=1}^N {\cal N} \left( \vartheta_i(\state^{(i)}) | {\bm \mu}_N^{\bm \vartheta},{\bm \Sigma}_N^{\bm \vartheta}  \right) ,
\label{GaussianCostFunctionRelabeling}
\end{equation}
with 
\begin{eqnarray}
{\bm \mu}_N^{\bm \vartheta} & = & \frac{1}{N} \sum_{i=1}^N \vartheta_i(\state^{(i)}) ,\\
{\bm \Sigma}_N^{\bm \vartheta}  &= & \frac{1}{N} \sum_{i=1}^N (\vartheta_i(\state^{(i)})-{\bm \mu}_N^{\bm \vartheta} )(\vartheta_i(\state^{(i)})-{\bm \mu}_N^{\bm \vartheta} )^T .
\end{eqnarray}
The Gaussian cost function in (\ref{GaussianCostFunctionRelabeling}) imposes the idea that one wants a relabeled sample to be the most Gaussian possible among its permutations ${\bm \vartheta}(\utilde{\state})$, ${\bm \vartheta} \in {\cal P}^N$ , in order for ${\bm \vartheta}(\utilde{\state})$ to look as unimodal as possible.

However, this technique is particularly costly since it involves a combinatorial optimization over  ${\cal P}^N$, which is unfeasible in practice: here the posterior is defined on $\mathbb{R}^{3K}$  and ${\cal P}$ is the group  formed by the permutations of $K$ elements, ${\cal P}^N$ has cardinality $(K!)^N$. As a consequence, in this work, we use the online version of this algorithm proposed in \cite{CeleuxOnline98} and having a final cost of $N (K!)$. To avoid the use of the resampling in order to get this set of unweighted particles, $\utilde{\state}$, we propose an adaptation of this algorithm, described in Algo. \ref{PostProcessingRelabelingAlgo}, in order to be able to use directly the set of weighted particles provided by the SMC sampler.

\begin{algorithm}[h]  
\caption{Online post-processing relabeling algorithm}
\label{PostProcessingRelabelingAlgo}
 \begin{algorithmic}[1]
   \small 
   \STATE \textbf{Input:} Set of weighted particles from the last iteration of the SMC sampler $\left\{\state_{k^*,T}^{(m)},\NormISWeight_{k^*,T}\right\}_{m=1}^N$
   \STATE Sort the particle by descending order of their associated weights
   \STATE Set ${\bm \mu}_1=\state_{k^*,T}^{(1)}$ and $\state_{relabel}^{(1)}=\state_{k^*,T}^{(1)}$  and initialize ${\bm \Sigma}_{1}$
   \FOR {$n=2, \ldots, N$}
   \STATE  Find $$ \vartheta_n = \argmax_{\vartheta \in {\cal P}} {\cal N} \left( \vartheta(\state_{k^*,T}^{(n)}) | {\bm \mu}_{n-1},{\bm \Sigma}_{n-1}  \right)$$
   \STATE Set $\state_{relabel}^{(n)} = \vartheta_{n}(\state_{k^*,T}^{(n)})$
   \STATE Set $\alpha^{(i)}=\NormISWeight_{k^*,T}^{(i)}\left[\sum_{j=1}^n \NormISWeight_{k^*,T}^{(j)} \right]^{-1}$ for $i=1,\ldots,n$ and compute:
     \begin{align*}
     \begin{split}
   {\bm \mu}_{n} & =  \sum_{i=1}^{n} \alpha^{(i)} \state_{relabel}^{(i)} \\
{\bm \Sigma}_{n}   &=  \sum_{i=1}^n \alpha^{(i)} (\state_{relabel}^{(i)}-{\bm \mu}_{n} )(\state_{relabel}^{(i)}-{\bm \mu}_{n} )^T
\end{split}
     \end{align*}

         \ENDFOR
        \STATE Use the relabeled collection of weighted particles $\left\{\state_{relabel}^{(m)},\NormISWeight_{k^*,T}^{(m)}\right\}_{m=1}^N$ to compute point estimate, e.g. MMSE:
        $$\widehat{\state}_{k^*}=\sum_{i=1}^N \NormISWeight_{k^*,T}^{(i)} \state_{relabel}^{(i)} $$
\end{algorithmic}
\end{algorithm}

\section{Posterior Cram\'er-Rao bound for multiple source localization}
\label{Sec:PCRB}
In this section, we derive the posterior Cram\'er-Rao bound (PCRB) as an estimation benchmark for the parameters. We will thus assume in this setting that we condition on the number of sources. This PCRB will thus provide a theoretical performance limit for the Bayesian estimator of the locations as well as the transmitted powers of the $K$ sources given the observations, ${\bm z}$, obtained at the fusion center. Let us remark that in \cite{Ozdemir2009}, the authors have derived the Cram\'er-Rao bound for the single source problem with quantized data and imperfect channel between the sensors and the fusion center. Here, we propose to generalize this result by considering $\state_K$ as a random variable (Bayesian framework which leads to the posterior CRB) and $\state_K$ composed of $K$ multiple sources.

Indeed, the PCRB gives a lower bound for the error covariance matrix 
\cite{VanTrees68}:
\begin{equation}
\Expec\left[\left(\hat{\state}_K({\bm z})-\state_K\right) \left(\hat{\state}_K({\bm z})-\state_K\right)^T \right] \geq {\bm J}^{-1} , 
\end{equation}
where ${\bm J}$ is the $3K \times 3K$ Fisher information matrix (FIM)
\begin{eqnarray}
{\bm J}&=&\Expec\left[ \nabla_\state \log p({\bm z},\state_K|{\cal M}_K)  \nabla_\state^T \log p({\bm z},\state_K|{\cal M}_K)   \right] \nonumber \\
&= & - \Expec \left[ \Delta_\state^\state \log p({\bm z},\state_K|{\cal M}_K)    \right] ,
\label{PosteriorFIM}
\end{eqnarray}
where $\Delta_\state^\state := \nabla_\state \nabla_\state^T$ is the second derivative operator and $\nabla_\state$ is the gradient operator with respect to $\state$.

Using the fact that $p({\bm z},\state_K|{\cal M}_K)=p({\bm z}|\state_K,{\cal M}_K)p(\state_K|{\cal M}_K)$, the expression of the FIM in (\eqref{PosteriorFIM}) can be expressed as:
\begin{equation}
{\bm J}=- \Expec \left[ \Delta_\state^\state \log p({\bm z}|\state_K,{\cal M}_K)    \right] - \Expec \left[ \Delta_\state^\state \log p(\state_K|{\cal M}_K)    \right] = {\bm J}_d + {\bm J}_p ,
\label{DecompoFIM}
\end{equation}
where $ {\bm J}_p$ represents the \textit{a priori} information and  ${\bm J}_d$ is the ``standard'' FIM (used in the derivation of the CRB) averaged over the prior of the different location and power of the $K$ sources:
\begin{equation}
{\bm J}_d=\int_{\Theta_k} {\bm J}_d(\state_K) p(\state_K|{\cal M}_K)  d\state_K .
\label{PosteriorFIMData}
\end{equation}

As demonstrated in the Appendix, this standard FIM is defined for this problem as follows:
\begin{equation}
{\bm J}_d(\state_K)=\sum_{i=1}^N \sum_{j=0}^{L-1} \frac{\nabla_\state  p({z}_i=j|\state_K,{\cal M}_K)  \nabla_\state^T  p({z}_i=j|\state_K,{\cal M}_K)}{p({z}_i=j|\state_K,{\cal M}_K)} ,
\label{StandardFIMData}
\end{equation}
with the gradient operator given by:
\begin{equation}
\nabla_\state =\begin{bmatrix}
\frac{\partial}{\partial P_1} & \frac{\partial}{\partial x_1} & \frac{\partial}{\partial y_1} & \cdots & \frac{\partial}{\partial P_K} & \frac{\partial}{\partial x_K} & \frac{\partial}{\partial y_K} ,
\end{bmatrix}^T .
\end{equation}
Using (\ref{LikelihoodTargetSingleObs}), the gradient term in (\ref{StandardFIMData}) is expressed as:
\begin{equation}
	\nabla_\state p(z_i=j|\state_K,{\cal M}_K)  =  \sum_{l=0}^{L-1} p(z_i=j | b_i=l) \nabla_\state p(b_i=l | \state_K,{\cal M}_K) ,
\end{equation}
in which for $k=1,\ldots,K$ produces:
\begin{eqnarray}
\frac{\partial p(b_i=l | \state_K,{\cal M}_K)}{\partial P_k} & = & \left(\frac{d_0}{d_{i,k}} \right)^{n/2} \frac{\rho_{i,l}}{2 \sqrt{2 \pi \sigma ^2 P_k}} , \nonumber \\
\frac{\partial p(b_i=l | \state_K,{\cal M}_K)}{\partial x_k} & = & \left(\frac{d_0}{d_{i,k}} \right)^{n/2} \frac{n P_k^{1/2} d_{i,k}^{-2} \rho_{i,l} (p_{x,i}-x_k)}{2 \sqrt{2 \pi \sigma ^2}} , \\
\frac{\partial p(b_i=l | \state_K,{\cal M}_K)}{\partial y_k} & = & \left(\frac{d_0}{d_{i,k}} \right)^{n/2} \frac{n P_k^{1/2} d_{i,k}^{-2} \rho_{i,l} (p_{y,i}-y_k)}{2 \sqrt{2 \pi \sigma ^2}} , \nonumber
\end{eqnarray}
and 
\begin{equation}
\rho_{i,l}=\left(e^{-\frac{(\lambda_{i,l}-a_i)^2}{2 \sigma^2}}- e^{-\frac{(\lambda_{i,l+1}-a_i)^2}{2 \sigma^2}}\right) .
\end{equation}
Although an analytical expression for ${\bm J}_d(\state_K)$ has been derived, in order to obtain ${\bm J}_d$ involved in the computation of the FIM defined in (\ref{DecompoFIM}), we need to resort to some numerical  techniques for the approximation of the integral that defines this quantity in (\ref{PosteriorFIMData}). The procedure we use is a simple Monte-Carlo integration:
\begin{enumerate}
\item Draw $N_{MC}$ realization of the state from the prior: $\left\{ \state_K^{i} \right\}_{i=1}^{N_{MC}} \sim p(\state_K)$.
\item Approximate the quantity of interest by:
\begin{equation}
{\bm J}_d\approx \frac{1}{N_{MC}} \sum_{i=1}^{N_{MC}} {\bm J}_d(\state_K^{i}).
\end{equation}
\end{enumerate}

Finally, the second term representing the \textit{a priori} information in (\ref{DecompoFIM}) is a $3K\times3K$ matrix defined as:
\begin{equation}
 {\bm J}_p=\begin{bmatrix}
 \xi & & & & \\
      & {\bm \Sigma}_p^{-1} & & 0 & \\
      & & \ddots & & \\
      & 0 & & \xi & \\
      &  & & & {\bm \Sigma}_p^{-1}
 \end{bmatrix} ,
\end{equation}
with $\xi=\frac{a(a+1)(a+3)}{b^2}  $ - \textit{Proof:} See the Appendix.

\section{Numerical Simulations}

In all the experiments, we consider a signal decay exponent and a reference distance as $n=2$ and $d_0=1$ respectively. The ROI is a $100m\times 100 m$ field in which 100 sensors are deployed in a grid where the location of each sensor is assumed to be known. 
The thresholds of the $M$-bit quantizer defined in (\ref{QuantizationChap3}) are the same for each sensor and are equally spaced between $\lambda_{i,1}=0$ and $\lambda_{i,L-1}=22$, $\forall i=1,\ldots,100$. The hyper-parameters in the prior distribution of the  transmitted power of each source defined in (\ref{PriorTargetChap3Details}) are $a=50$ and $b=2.5\times 10^5$. An uniform distribution is used as the prior over the collection of models, i.e. $\forall k\in \left\{1,\ldots,K_{\max} \right\}$ we have $p(\model_k)=1/K_{\max}$. 
All the results have been obtained by using $\NbMCMCmove=5$ in the  MWG (summarized in Algo. \ref{Algo_AMWG}) used in the SMC sampler as forward kernel. In order to illustrate the benefit of using the SMC sampler, we have adapted to the problem considered in this paper the importance sampler (IS) that has been proposed for a single source localization in \cite{Masazade2010}. For each model $\model_k$, this IS algorithm simply consists in sampling $N_{IS}$ particles from the prior distribution in (\ref{PriorTargetChap3Details}) and in assigning  to each of these particles an weights which is proportional to the likelihood given in (\ref{CompleteLikelihood}). In order to have a fair comparison between both algorithms since the proposed SMC sampler adapts the number of iterations on-line using the procedure described in Section \ref{ProcedureAdaptiveForward}, the number of particles used in the IS algorithm is set to $N_{IS}=TN$, with $T$  the total number of SMC iterations averaged over multiple runs for the configuration under study. As a consequence, the complexity of these two algorithms is equivalent since the number of particles generated for all the results described below is the same for both schemes.

\subsection{Accuracy of the estimators}

\begin{figure}[!htb]
\centering
\subfloat[]{
\includegraphics[width=0.5\textwidth]{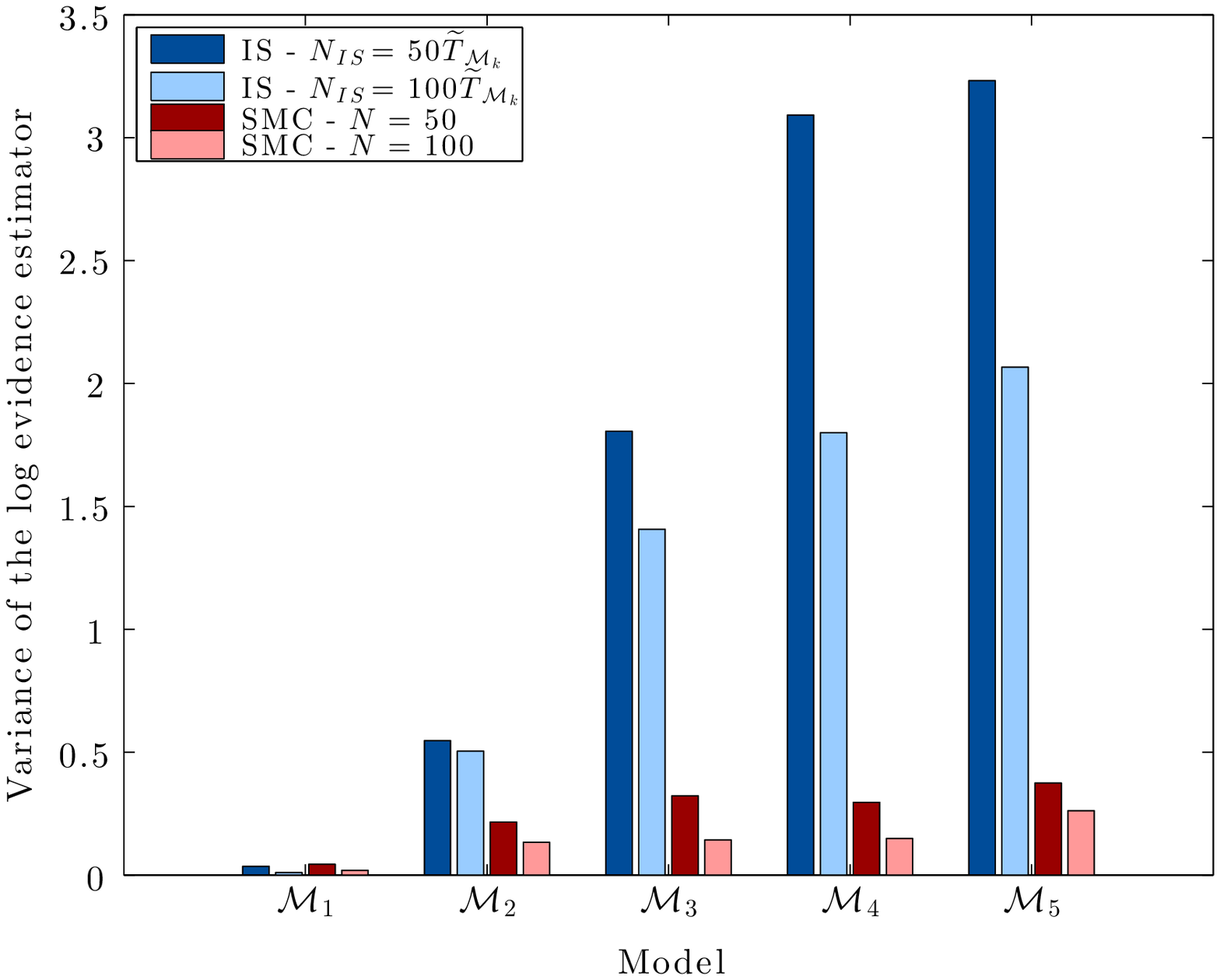} \label{fig:var_logevidence}}
\subfloat[]{
\includegraphics[width=0.5\textwidth]{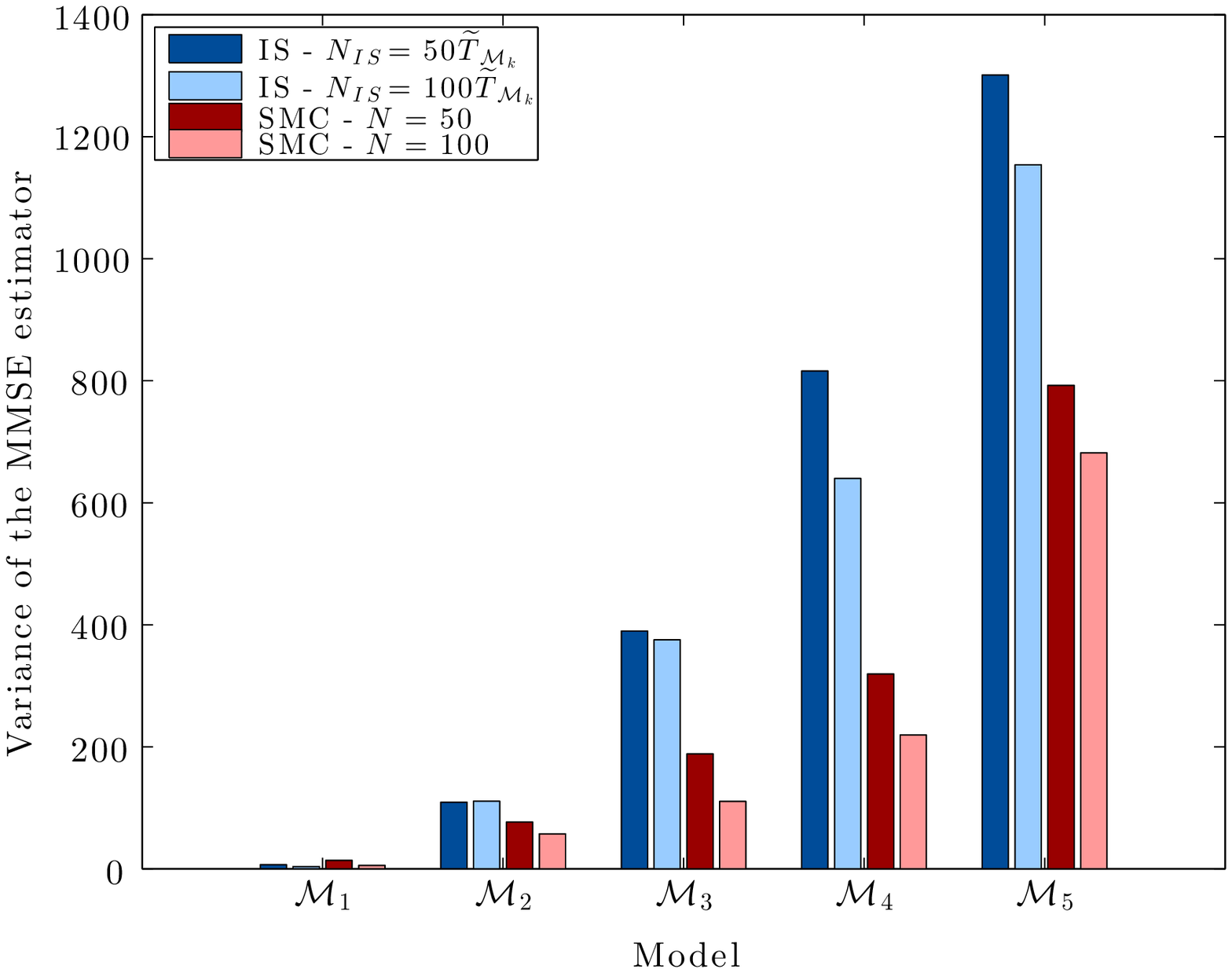} \label{fig:var_MMSE}}
\caption{Comparison of the variance of both the log evidence, $\log p({\bm z}|{\cal M}_k)$, estimator in (a) and MMSE estimator in (b) provided by the proposed SMC sampler and the IS algorithm - The parameter of the adaptive cooling schedule is set as $\overline{\CESS}=0.9 N$ leading to the following average number of iterations $\widetilde{T}_{{\cal M}_1}=8$, $\widetilde{T}_{{\cal M}_2}=12$, $\widetilde{T}_{{\cal M}_3}=15$, $\widetilde{T}_{{\cal M}_4}=17$ and $\widetilde{T}_{{\cal M}_5}=18$ for both $N=50$ or $N=100$ (Number of quantization levels: $L=4$ and $\sigma^2=1$)}
\label{fig:VarianceLogEvidence}
\end{figure}

\begin{figure}[!htb]
\centering
\includegraphics[width=0.5\textwidth]{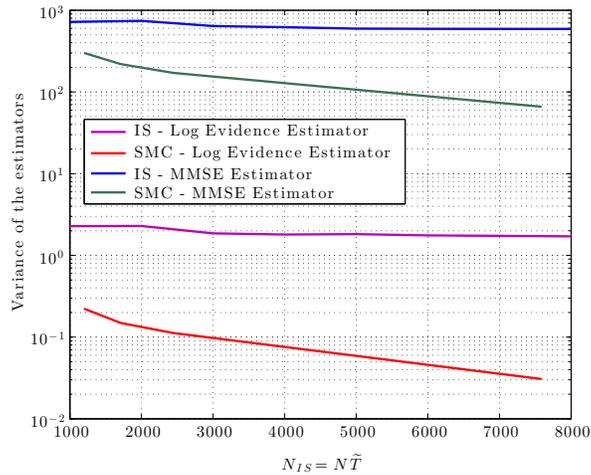} 
\caption{Evolution of the variance of the estimators of both the log evidence and the MMSE provided by the IS and the SMC for model ${\cal M}_4$. For the SMC, the results have been obtained with $N=100$ particles by varying the  $\overline{\CESS}$ thus leading to a different average number of iterations, $\widetilde{T}$.  (Number of quantization levels: $L=4$ and $\sigma^2=1$ )}
\label{fig:EvolutionVariance}
\end{figure}

We first study the robustness and the accuracy of the estimators of the two main quantities of interest: model posterior probabilities and the MMSE of the parameters under each model. In order to perform this analysis, both schemes have been run 100 times on the same realization of observations from a scenario with 4 sources. From the theory, we know that both IS and SMC algorithms provides, whatever the number of particles used, an unbiased estimate of  $\log p({\bm z}|{\cal M}_k)$ which corresponds as discussed in (\ref{PosteriorModelChap3}) to the only unknown quantity in the model posterior distribution. However, as depicted in Fig. \ref{fig:var_logevidence}, the variance of the estimator of this quantity obtained from these two algorithms is significantly different. If both algorithms perform similarly for one source, the SMC sampler outperforms significantly the IS algorithm as the number of sources increases. The same remark holds for the variance of the MMSE estimator shown in Fig. \ref{fig:var_MMSE} which is quite remarkable since the MMSE with the SMC sampler is only computed with $N$ particles instead of $N\widetilde{T}$ with the IS.  The same remarks hold even for an increasing number of particles as illustrated in Fig. \ref{fig:EvolutionVariance}. To understand these results, we present in Table \ref{table:ESS} the effective sample size (ESS), defined in (\ref{Eq_SMC_ESS}) which represents the number of particles that will really contribute to the final estimator. We can see from these ESS results that the IS algorithm completely collapses when the dimension of the model (i.e. the number of sources in the ROI) increases. As an example, for the model ${\cal M}_4$, only 3-4 particles over the  $N_{IS}=N\widetilde{T}=1700$ will really contribute to the MMSE estimator by having a non-negligible importance weights. Conversely, the ESS from the SMC sampler is quite stable with the dimension of the model. All these results clearly illustrate the significant gain provided by the proposed SMC sampler and thus the benefit of  gradually introducing the likelihood information through the successive iterations of the algorithm.

\begin{table}[htb]                            
\centering                                   
\begin{tabular}{c|c|c|c|c|c}                 
 & ${\cal M}_1$ & ${\cal M}_2$ & ${\cal M}_3$ & ${\cal M}_4$ & ${\cal M}_5$ \\
\hline                                     
IS & 0.0836 & 0.0082 & 0.0030 & 0.0019 & 0.0018 \\
\hline                                       
SMC & 0.6180 & 0.6290 & 0.6297 & 0.6276 & 0.6319 \\
\hline                                       
\end{tabular}                                
\caption{Comparison of the average Effective sample size defined in (\ref{Eq_SMC_ESS}) and scaled by the number of particles for both SMC with $N=100$ and IS algorithms for the different models}                     
\label{table:ESS}                   
\end{table}

\begin{figure}[!htb]
\centering
\subfloat[Posterior marginal before relabeling]{
\includegraphics[width=0.5\textwidth]{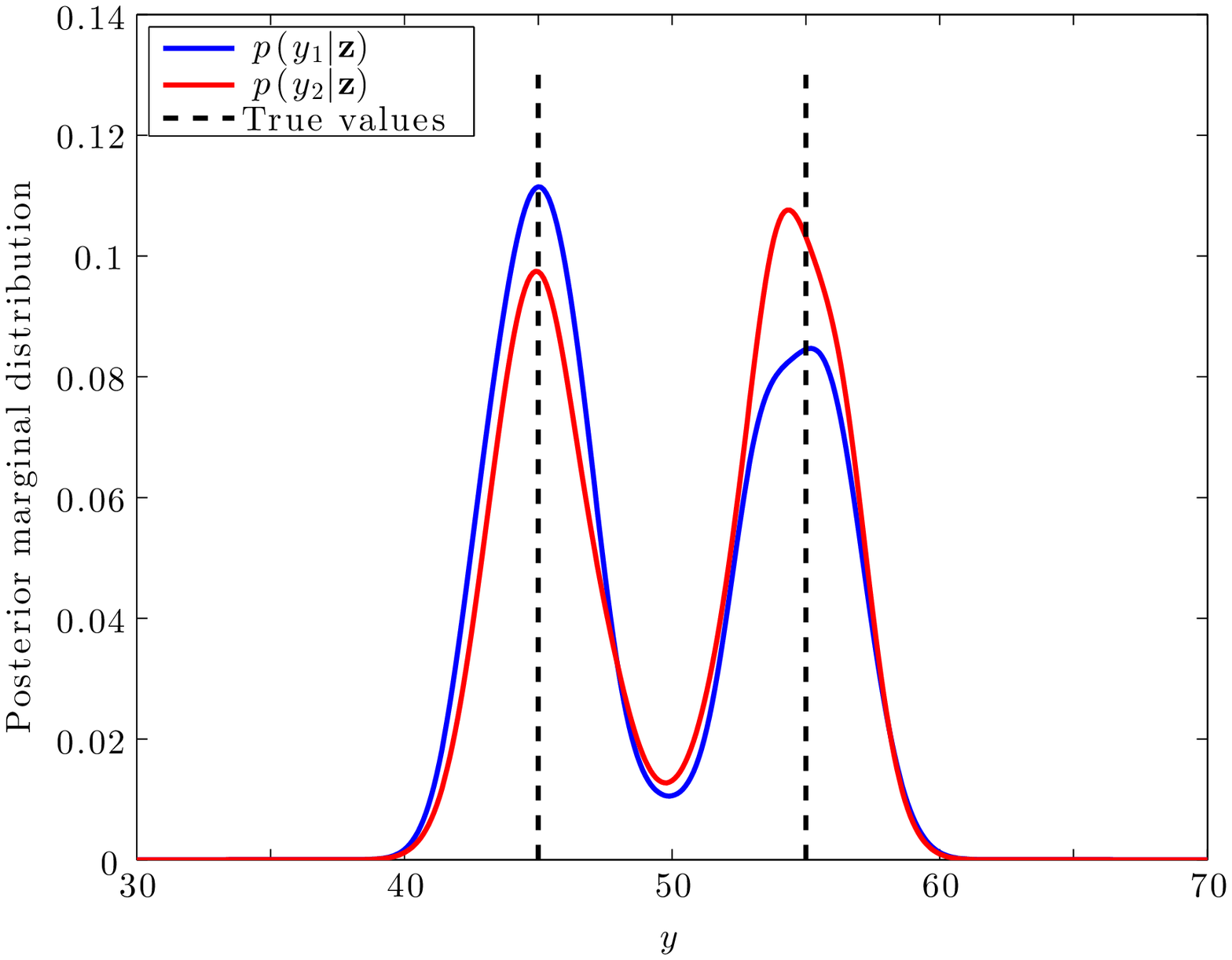} }
\subfloat[Posterior marginal  after relabeling]{
\includegraphics[width=0.5\textwidth]{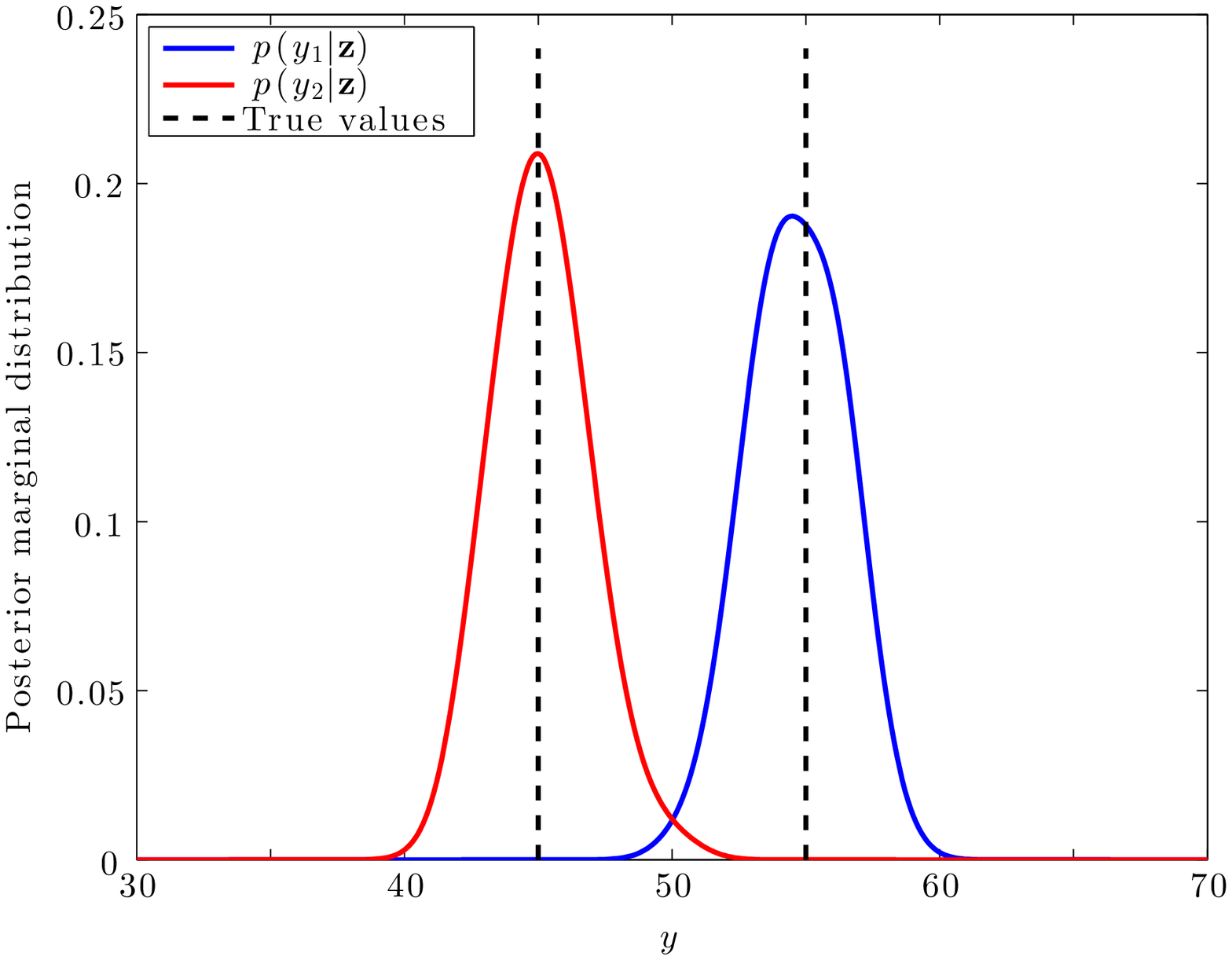} }\\
\subfloat[Model Posterior]{
\includegraphics[width=0.5\textwidth]{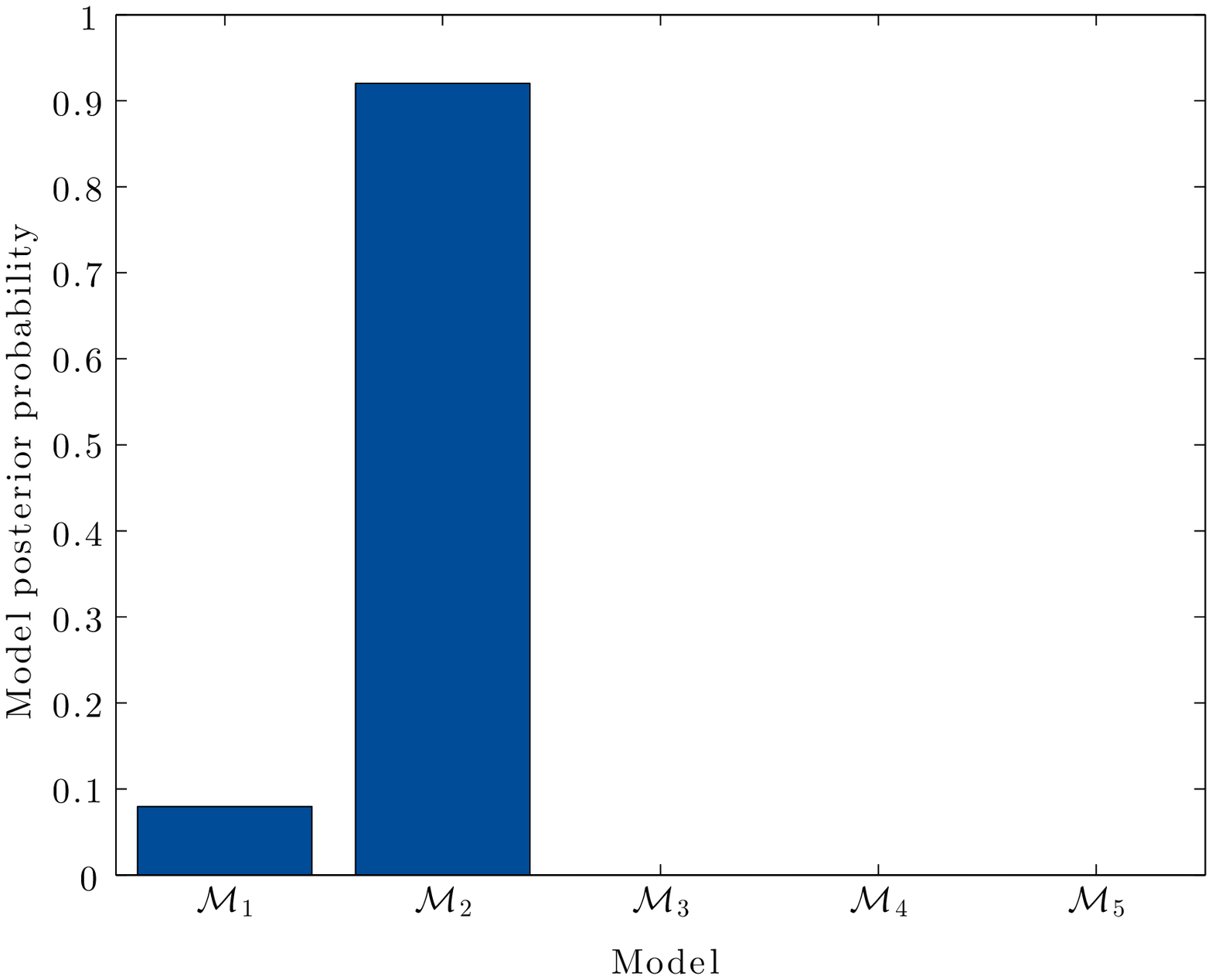} }
\caption{Illustration of the relabeling and the model posterior obtained with the proposed SMC sampler ($N=200$ and  $\overline{\CESS}=0.9N$) in a scenario with two targets located at $[50,45]$ and $[50,55]$ (Number of quantization levels: $L=8$ and $\sigma^2=0.5$ )  }
\label{fig:MultimodalPosteriorCase1}
\end{figure}

Let us now illustrate with a 2 targets scenario, the label switching problem discussed in Section \ref{ProposedLabelSwitching} and the importance of having a relabeling algorithm in order to provide a point estimate. In Fig. \ref{fig:MultimodalPosteriorCase1}, we present the marginal posterior distribution obtained with the proposed SMC sampler. We can first remark that the algorithm is clearly able to capture the multimodality of each marginals. However, if the MMSE is directly computed from this approximation, the estimated $y$-coordinate for both targets will be approximately 50 instead of 55 and 45. The proposed relabeling algorithm described in Algo. \ref{PostProcessingRelabelingAlgo} allows to isolate both modes by finding the best permutations for all particles. The MMSE estimate by taking the particle system after relabeling will therefore provide an accurate point estimate close to the truth. Moreover from the estimates of the posterior distribution in this figure, we can remark that the algorithm is able to detect that there are 2 targets in the scene. 

\begin{figure}[!htb]
\centering
\subfloat[Posterior marginal before relabeling]{
\includegraphics[width=0.5\textwidth]{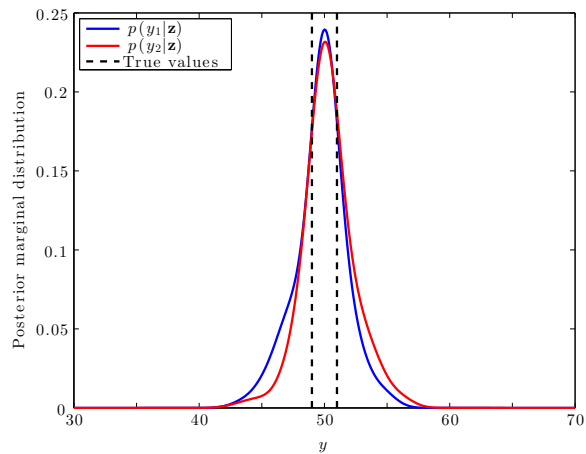} }
\subfloat[Posterior marginal after relabeling]{
\includegraphics[width=0.5\textwidth]{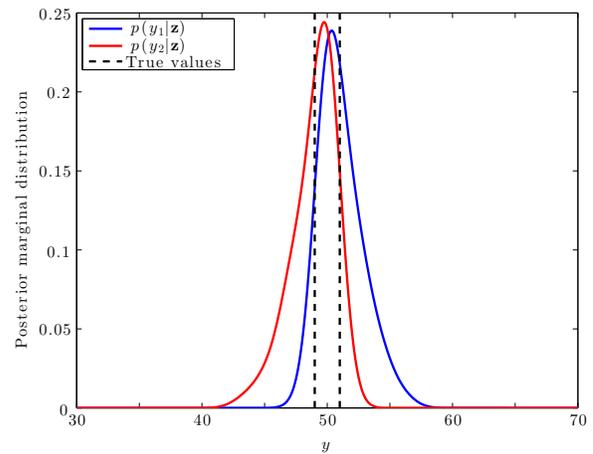} }\\
\subfloat[Model Posterior]{
\includegraphics[width=0.5\textwidth]{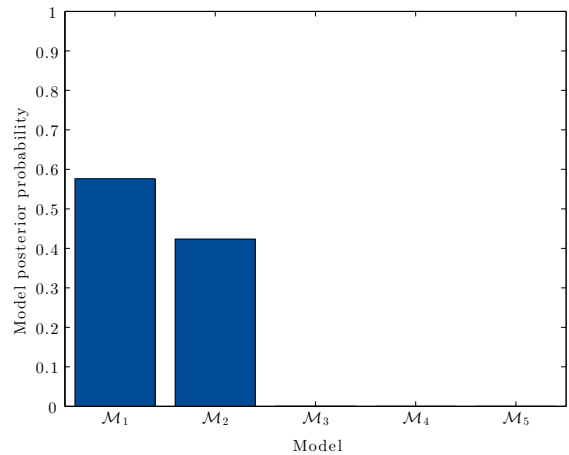} }
\caption{Illustration of the relabeling and the model posterior obtained with the proposed SMC sampler ($N=200$ and  $\overline{\CESS}=0.9N$) in a scenario with two close targets located at $[50,49]$ and $[50,51]$ (Number of quantization levels: $L=8$ and $\sigma^2=0.5$ ) }
\label{fig:MultimodalPosteriorCase2}
\end{figure}

Let us now illustrate with Fig. \ref{fig:MultimodalPosteriorCase2}, the challenging problem of having two targets of interest that are placed very close to each other ($[50,49]$ and $[50,51]$). We remark that from the model posterior probabilities that the algorithm is able to provide the uncertainty arising from this scenario between the model with 1 and 2 targets. The ability of the algorithm to distinguish two close targets will clear be a function of many parameters of the sensor network: distance between sensors, variance of the observation noise, number of quantization levels as well as the quality of the wireless channel between the sensors and the fusion center. The approximation of the posterior marginal distributions obtained from the SMC sampler under model ${\cal M}_2$ are both unimodal owing to the relatively small distance between the 2 sources. Once again in this case, we can see the benefit of using the relabeling algorithm for computing the final MMSE point estimator.

Next, we compare the performances of the proposed algorithm when 4 sources are in the ROI.  In order to obtain the following results, 100 realizations of the four sources and associated observations have been drawn from the prior and likelihood defined in Section \ref{ProblemFormulationSec}. Table \ref{table:ModelSelection} illustrates the ability of the proposed to detect the correct number of targets. The correct number  of target ( i.e. Model ${\cal M}_4$) is chosen more often with the proposed SMC sampler than the IS algorithm over the 100 scenarios.  Even if both provides an unbiased estimate of the evidence of each model, the significant lower variance of the estimator provided by the SMC sampler, which was illustrated previously in Fig. \ref{fig:VarianceLogEvidence}, allow to have a more efficient and accurate model decision step.

\begin{table}[htb]                            
\centering                                   
\begin{tabular}{c|c|c|c|c|c}                 
 & ${\cal M}_1$ & ${\cal M}_2$ & ${\cal M}_3$ & ${\cal M}_4$ & ${\cal M}_5$ \\
\hline                                     
IS & 0 & 0 & 13 & 85 & 2 \\
\hline                                                  
SMC & 0 & 0 & 3 & 96 & 1  \\
\hline                                       
\end{tabular}                                
\caption{Comparison of the number of times that each model has been selected from the estimated model posterior probabilities given by both the proposed SMC sampler the IS algorithm under 100 realizations ($N=100$, $\overline{\CESS}=0.9N$, $L=4$ and $\sigma^2=1$)}                     
\label{table:ModelSelection}                   
\end{table}  

\subsection{Localization Performance and PCRB}

\begin{figure}[!htb]
\centering
\subfloat[$\sigma^2=1$]{
\includegraphics[width=0.5\textwidth]{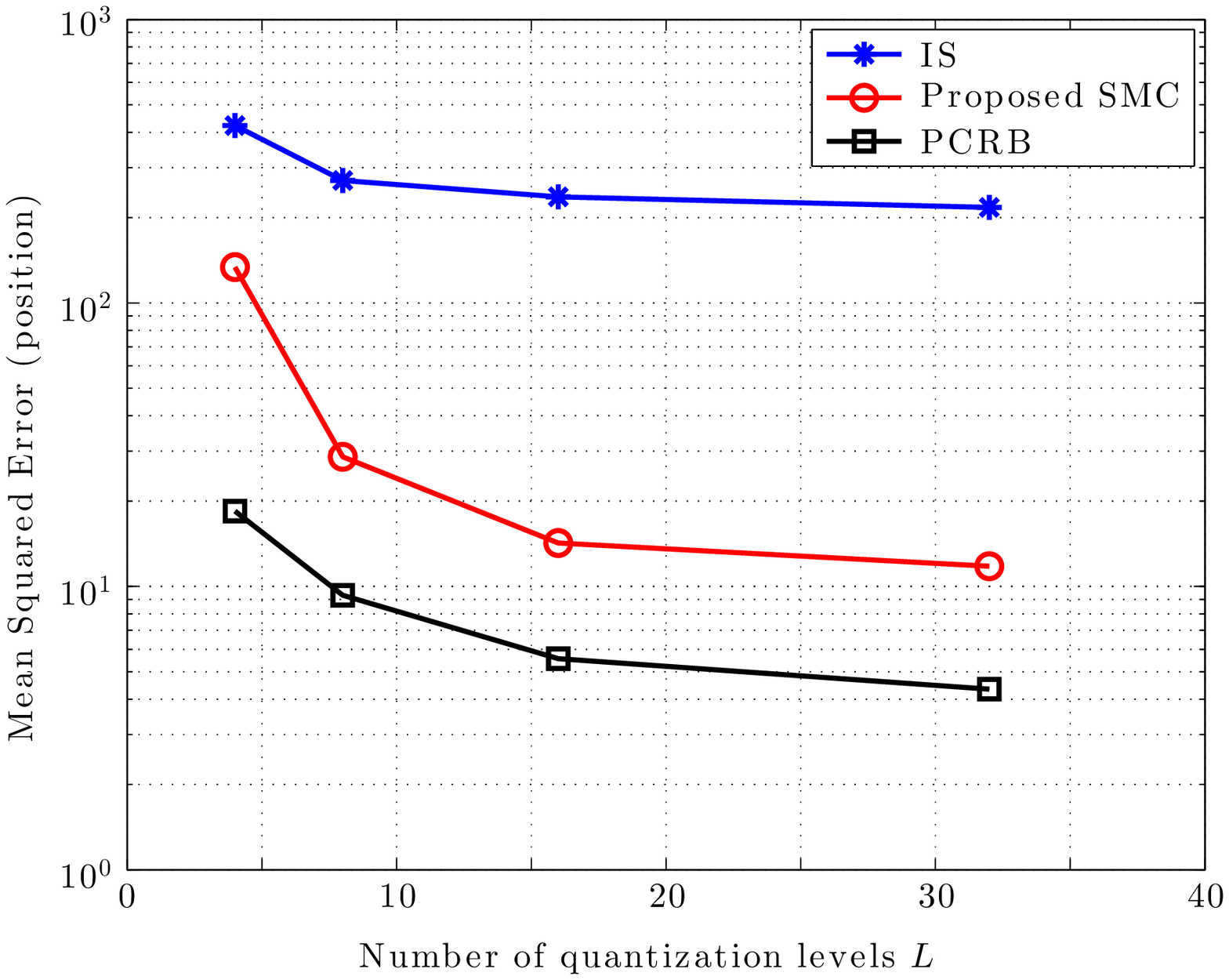} }
\subfloat[$\sigma^2=0.05$]{
\includegraphics[width=0.5\textwidth]{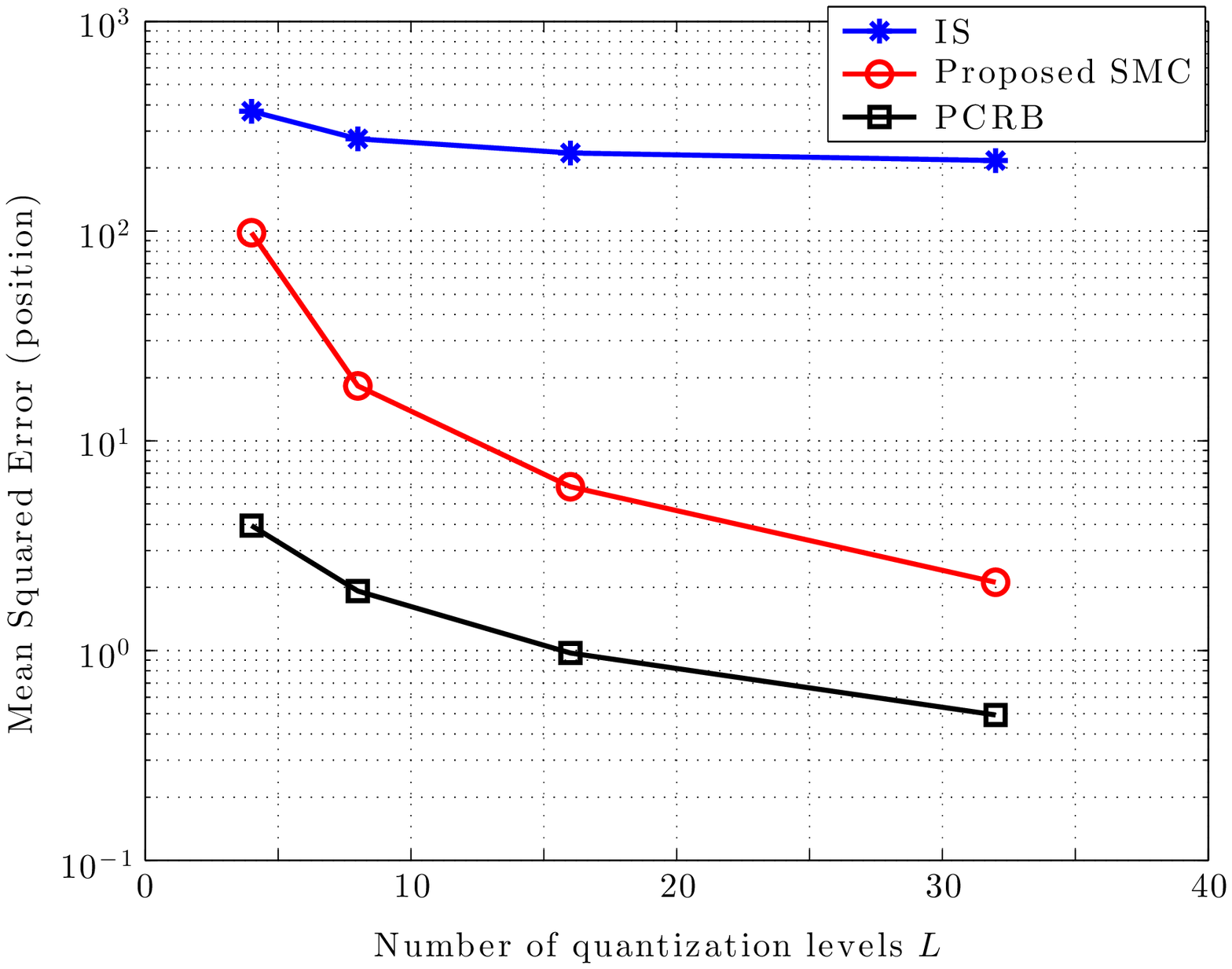} }
\caption{Evolution of the mean squared error for the source locations as a function of the number of quantization levels $L$ with two different values of the measurement noise $\sigma^2$ ($N=100$ and $\overline{\CESS}=0.9N$)}
\label{fig:MSE}
\end{figure}

\begin{figure}[!htb]
\centering
\subfloat[$\sigma^2=1$]{
\includegraphics[width=0.5\textwidth]{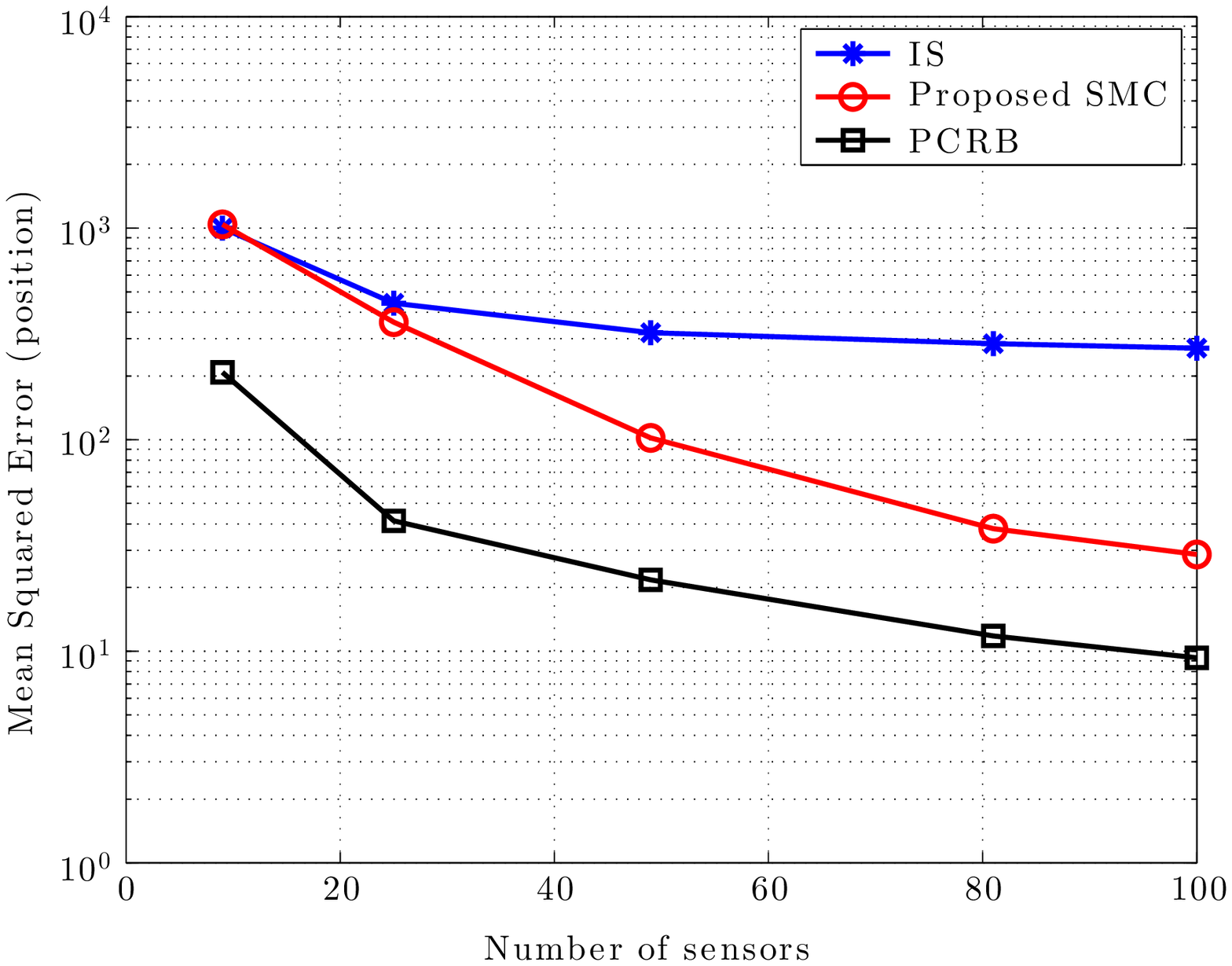} }
\subfloat[$\sigma^2=0.05$]{
\includegraphics[width=0.5\textwidth]{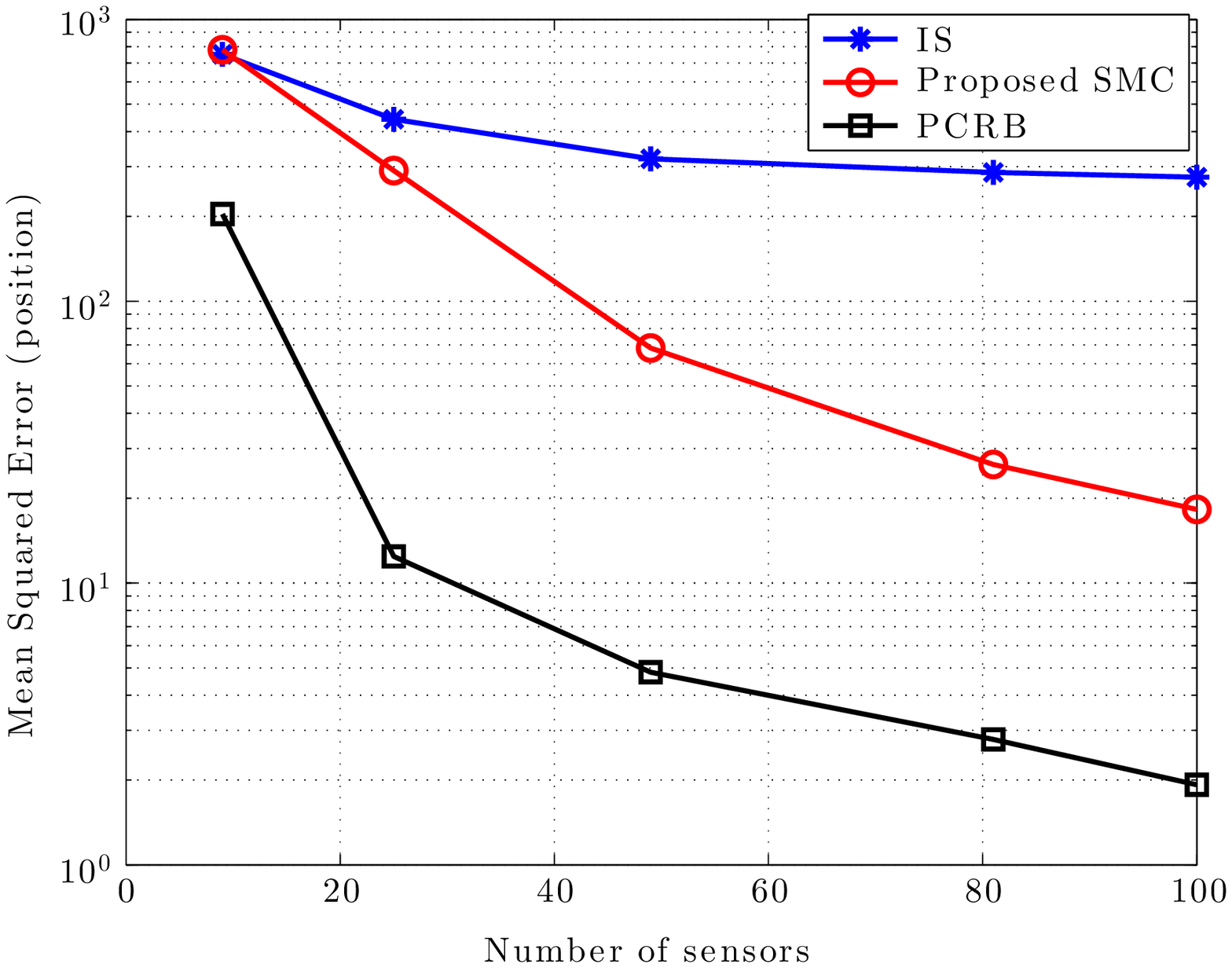} }
\caption{Evolution of the mean squared error for the source locations as a function of the number of sensors in the ROI  with two different values of the measurement noise $\sigma^2$ (Number of quantization levels: $L=8$ - $N=100$ and $\overline{\CESS}=0.9N$)}
\label{fig:MSENbSensors}
\end{figure}

In Fig. \ref{fig:MSE} the performance of the proposed SMC sampler (and the IS algorithm) in term of the mean squared error between point estimate $\hat{\state}_p$ of the algorithm  and the true location ${\state}_p$ of the four sources:
\begin{equation}
MSE=\text{trace} \left\{ \Expec \left[ (\hat{\state}_p- {\state}_p) (\hat{\state}_p - {\state}_p)^T \right]\right\} 
\end{equation}
with $\hat{\state}_p=\begin{bmatrix}
\hat{x}_1 & \hat{y}_1 & \cdots & \hat{x}_4 & \hat{y}_4
\end{bmatrix}^T$ and $\state_p=\begin{bmatrix}
{x}_1 & {y}_1& \cdots & {x}_4 & {y}_4
\end{bmatrix}^T$ represents the estimated (by using the relabeling algorithm) and the true location of the four targets, respectively. 
We also plot the associated PCRB that we have derived in Section \ref{Sec:PCRB}.  In order to obtain the results we use 100 realizations (of the different source characteristics and associated observations by avoiding the case in which two targets are very close). The results depicted in Figures \ref{fig:MSE} and \ref{fig:MSENbSensors} clearly demonstrate the good localization performance of the proposed algorithm and the significant gain compared to the IS algorithm which completely fails to localize four targets. As expected, the accuracy on the localization improves with the increase of either the number of sensors or the number of quantization levels as well as with the decrease of the measurement noise variance. 

\section{Conclusion}
In this paper, we addressed the problem of localizing an unknown number of energy emitting sources in wireless sensor networks with quantized data. We provided a generalization of recent existing works considering a single source. We  proposed a Bayesian solution for the joint estimation of the unknown number of sources as well as their associated parameters which is based on SMC sampler. Then, we derived the posterior Cram\'er-Rao bound for the estimation of the characteristics of these multiple energy emitting sources. Numerical simulations clearly illustrated the ability of the proposed SMC sampler to perform this challenging joint estimation. The different experiments showed that the proposed scheme based on novel SMC sampler improves quite significantly the accuracy of the estimators  method that are required for model selection (i.e., the number of sources) and the estimation of the source characteristics compared to more classical importance sampling method.

\appendix[Proof of the Posterior Cram\'er-Rao bound]

\label{DemonstrationPCRB}

As presented in Section \ref{Sec:PCRB}, the Fisher information matrix (FIM) for the posterior Cram\'er-Rao bound (PCRB) can be decomposed as follows:
\begin{equation}
{\bm J}  =  \underbrace{\int_{\Theta_k} {\bm J}_d(\state_K) p(\state_K|{\cal M}_K)  d\state_K}_{{\bm J}_d}  + \underbrace{\Expec \left[ -\Delta_\state^\state \log p(\state_K|{\cal M}_K)    \right]}_{{\bm J}_p}
\end{equation}
where $\Delta_\state^\state := \nabla_\state \nabla_\state^T$ is the second derivative operator and $\nabla_\state$ is the gradient operator with respect to $\state$.

In this appendix, we derive respectively ${\bm J}_d(\state_K)$ and ${\bm J}_p$.

The information matrix ${\bm J}_d(\state_K)$ is defined as:
\begin{equation}
{\bm J}_d(\state_K)=-\Expec_{{\bm z}|\state_K} \left[ \Delta_\state^\state \log p({\bm z}|\state_K,{\cal M}_K)    \right] .
\label{Appen_FIMData}
\end{equation}
The first derivative of the log likelihood is given by:
\begin{eqnarray}
 \nabla_\state^T \log p({\bm z}|\state_K,{\cal M}_K) & = & \sum_{i=1}^N \nabla_\state^T \log p({z}_i|\state_K,{\cal M}_K) \nonumber \\
 & = & \sum_{i=1}^N \frac{\nabla_\state^T  p({z}_i|\state_K,{\cal M}_K)}{p({z}_i|\state_K,{\cal M}_K)} ,
\end{eqnarray}
Therefore, the second derivative can be written as:
\begin{eqnarray}
\Delta_\state^\state \log p({\bm z}|\state_K,{\cal M}_K) & = & \nabla_\state \nabla_\state^T \log p({\bm z}|\state_K,{\cal M}_K) \nonumber \\
 & = & \sum_{i=1}^N \frac{\nabla_\state \nabla_\state^T  p({z}_i|\state_K,{\cal M}_K)}{p({z}_i|\state_K,{\cal M}_K)} \nonumber\\
 && -  \frac{\nabla_\state  p({z}_i|\state_K,{\cal M}_K) \nabla_\state^T  p({z}_i|\state_K,{\cal M}_K)}{p({z}_i|\state_K,{\cal M}_K)^2} .
\end{eqnarray}
To obtain (\ref{Appen_FIMData}), we now take the negative expectation of this second derivative with respect to $p({z}_i|\state_K,{\cal M}_k=K) $:
\begin{eqnarray}
{\bm J}_d(\state_K) & =&  \sum_{i=1}^N \sum_{j=0}^{L-1} p({z}_i=j|\state_K,{\cal M}_K) \left\{ - \frac{\nabla_\state \nabla_\state^T  p({z}_i=j|\state_K,{\cal M}_K)}{p({z}_i=j|\state_K,{\cal M}_K)}\right.\nonumber \\
&& \left. + \frac{\nabla_\state  p({z}_i=j|\state_K,{\cal M}_K) \nabla_\state^T  p({z}_i=j|\state_K,{\cal M}_K)}{p({z}_i=j|\state_K,{\cal M}_K)^2}\right\} \nonumber \\
& =&  \sum_{i=1}^N \sum_{j=0}^{L-1} \frac{\nabla_\state  p({z}_i=j|\state_K,{\cal M}_K) \nabla_\state^T  p({z}_i=j|\state_K,{\cal M}_K)}{p({z}_i=j|\state_K,{\cal M}_K)} \nonumber \\
&& - \nabla_\state \nabla_\state^T  p({z}_i=j|\state_K,{\cal M}_K) .
\end{eqnarray}
The second term is equal to 0 since:
\begin{eqnarray}
\sum_{i=1}^N \sum_{j=0}^{L-1} \nabla_\state \nabla_\state^T  p({z}_i=j|\state_K,{\cal M}_K) & = & \sum_{i=1}^N  \nabla_\state \nabla_\state^T \underbrace{\sum_{j=0}^{L-1} p({z}_i=j|\state_K,{\cal M}_K)}_{=1} \nonumber \\
& = & 0 .
\end{eqnarray}
As a consequence, we finally obtain:
\begin{equation}
{\bm J}_d(\state_K) = \sum_{i=1}^N \sum_{j=0}^{L-1} \frac{\nabla_\state  p({z}_i=j|\state_K,{\cal M}_K) \nabla_\state^T  p({z}_i=j|\state_K,{\cal M}_K)}{p({z}_i=j|\state_K,{\cal M}_K)}  .
\end{equation}

Using (\ref{LikelihoodTargetSingleObs}), the gradient term  involved in this expression can be expressed as:
\begin{equation}
	\nabla_\state p(z_i=j|\state_K,{\cal M}_K)  =  \sum_{l=0}^{L-1} p(z_i=j | b_i=l) \nabla_\state p(b_i=l | \state_K,{\cal M}_K) ,
\end{equation}
with 
\begin{equation}
p(b_i=l | \state_K,{\cal M}_K)= Q\left( \frac{\lambda_{i,l}-a_i}{\sigma} \right) -Q\left( \frac{\lambda_{i,l+1}-a_i}{\sigma} \right) .
\end{equation}
As a consequence, since the $Q$- function is the complementary Gaussian cumulative distribution, we can easily remark that:
\begin{equation}
\nabla_\state p(b_i=l | \state_K,{\cal M}_K) = \frac{1}{\sqrt{2 \pi \sigma^2}} \underbrace{\left(e^{-\frac{(\lambda_{i,l}-a_i)^2}{2 \sigma^2}}- e^{-\frac{(\lambda_{i,l+1}-a_i)^2}{2 \sigma^2}}\right)}_{\rho_{i,l}} \nabla_\state a_i .
\end{equation}

Finally from the definition of $a_i$ in (\ref{AmplitudeDefLoc}), we obtain, for $k=1,\ldots,K$:
\begin{eqnarray}
\frac{\partial p(b_i=l | \state_K,{\cal M}_K)}{\partial P_k} & = & \left(\frac{d_0}{d_{i,k}} \right)^{n/2} \frac{\rho_{i,l}}{2 \sqrt{2 \pi \sigma ^2 P_k}}  ,\nonumber \\
\frac{\partial p(b_i=l | \state_K,{\cal M}_K)}{\partial x_k} & = & \left(\frac{d_0}{d_{i,k}} \right)^{n/2} \frac{n P_k^{1/2} d_{i,k}^{-2} \rho_{i,l} (p_{x,i}-x_k)}{2 \sqrt{2 \pi \sigma ^2}} , \\
\frac{\partial p(b_i=l | \state_K,{\cal M}_K)}{\partial y_k} & = & \left(\frac{d_0}{d_{i,k}} \right)^{n/2} \frac{n P_k^{1/2} d_{i,k}^{-2} \rho_{i,l} (p_{y,i}-y_k)}{2 \sqrt{2 \pi \sigma ^2}} , \nonumber
\end{eqnarray}
which completes the analytical calculation of ${\bm J}_d(\state_K)$.

Finally, we derive the \textit{a priori} information matrix given by:
\begin{equation}
{\bm J}_p=\Expec \left[ -\Delta_\state^\state \log p(\state_K|{\cal M}_K)   \right] .
\end{equation}

From the prior distributions in (\ref{PriorTargetChap3}-\ref{PriorTargetChap3Details}), each target's location and power are independent and identically distributed. ${\bm J}_p$ will be therefore a $3K\times 3K$ block diagonal matrix with information associated to the location and the power defined respectively as:
\begin{equation}
\Expec \left[ -\Delta_{[x_k,y_k]^T}^{[x_k,y_k]^T} \log \Normal([x_k,y_k]^T| {\bm \mu}_p, {\bm \Sigma}_p)  \right]={\bm \Sigma}_p^{-1} ,
\end{equation}
and
\begin{eqnarray}
\Expec \left[ -\Delta_{P_k}^{P_k} \log  {\cal IG} (P_k|a,b) \right]&=&\Expec \left[ -\frac{\partial^2}{\partial P_k^2} \log \left\{\frac{b^a}{\Gamma(a)} P_k^{-a-1} \exp\left( -\frac{b}{P_k}\right)\right\}\right] \nonumber \\
& = & \Expec \left[ \frac{\partial^2}{\partial P_k^2} (a+1)\log(P_k) +\frac{b}{P_k}\right] \nonumber \\
& =&  \Expec \left[ 2bP_k^{-3} - (a+1)P_k^{-2}\right]\nonumber\\
& =&  2b\Expec \left[ P_k^{-3}\right] - (a+1)\Expec \left[ P_k^{-2}\right] .
\label{PrePriorInfoPower}
\end{eqnarray}
Let us now derive the two moments involved in this expression. We have, for $n>0$:
\begin{eqnarray}
\Expec \left[ P_{k}^{-n}\right] &=& \int_0^{+\infty} P_{k}^{-n}\frac{b^a}{\Gamma(a)} P_{k}^{-a-1} \exp\left( -\frac{b}{P_{k}}\right) dP_{k} \nonumber \\
& = & \frac{b^a}{\Gamma(a)} \int_0^{+\infty} P_{k}^{-(a+n)-1} \exp\left( -\frac{b}{P_{k}}\right) dP_{k}  \nonumber \\
& = & \frac{b^a}{\Gamma(a)} \frac{\Gamma(a+n)}{b^{a+n}} .
\end{eqnarray}
The last expression is obtained from the expression of the normalizing constant of an inverse-gamma distribution, ${\cal IG} (a+n,b)$. By using the equality of the Gamma function, $\Gamma(a+1)=a\Gamma(a)$, we obtain:
\begin{eqnarray}
\Expec\left[ P_k^{-2}\right] & =&  \frac{(a+1)a}{b^2} ,\\
\Expec\left[ P_k^{-3}\right] & =&  \frac{(a+2)(a+1)a}{b^3}.
\end{eqnarray}
By plugging these expressions in (\ref{PrePriorInfoPower}), the prior information for the power is given by:
\begin{equation}
\Expec \left[ -\Delta_{P_k}^{P_k} \log  {\cal IG} (P_k|a,b) \right] = \frac{a(a+1)(a+3)}{b^2}  =\xi ,
\end{equation}
leading to 
\begin{equation}
 {\bm J}_p=\begin{bmatrix}
 \xi & & & & \\
      & {\bm \Sigma}_p^{-1} & & 0 & \\
      & & \ddots & & \\
      & 0 & & \xi & \\
      &  & & & {\bm \Sigma}_p^{-1}
 \end{bmatrix} .
\end{equation}

\bibliographystyle{IEEEtran}
\bibliography{references,bibliography_Tracking,bibliography_Thesis,bibliography_Regression,bibliography_OtherChap2,referencesGWP}

\end{document}